\documentstyle[prl,aps,floats,epsf,color]{revtex}
\begin{document}
\baselineskip=12pt
\def\be{\begin{equation}}
\def\ee{\end{equation}}
\def\bea{\begin{eqnarray}}
\def\eea{\end{eqnarray}}
\def\orc{\Omega_{r_c}}
\def\om{\Omega_{\text{m}}}
\def\E{{\rm e}}
\def\bearst{\begin{eqnarray*}}
\def\eearst{\end{eqnarray*}}
\def\peleven{\parbox{11cm}}
\def\peffec{\peight{\bearst\eearst}\hfill\peleven}
\def\pspace{\peight{\bearst\eearst}\hfill}
\def\ptwelve{\parbox{12cm}}
\def\peight{\parbox{8mm}}
\twocolumn[\hsize\textwidth\columnwidth\hsize\csname@twocolumnfalse\endcsname

\title
{Is Thick Brane Model Consistent with Recent Observations?}

\author{ M. Sadegh Movahed$^{1}$ and Sima Ghassemi$^{2}$}
\address{$^{1}$Department of Physics, Shahid Beheshti University, Evin, Tehran 19839, Iran}
\address{$^{2}$ Institute for Studies in theoretical Physics and Mathematics, P.O.Box 19395-5531,Tehran,
Iran}

\vskip 1cm

 \maketitle

\begin{abstract}
There exist many observational evidences implying the expansion of
our universe is undergoing a late-time acceleration, the mechanism
of this acceleration is yet unknown. In the so-called thick brane
model this phenomena is attributed to the thickness of the brane
along the extra dimension. In this study we mainly rely to the
consistency of this model with most recent observational data
related to the background evolution. The new Supernova Type Ia
(SNIa) Gold sample and Supernova Legacy Survey (SNLS) data, the
position of the acoustic peak at the last scattering surface from
the Wilkinson Microwave Anisotropy Prob (WMAP) observations and the
baryon acoustic oscillation peak found in the Sloan Digital Sky
Survey (SDSS) are used to constrain the free parameter of the thick
codimension $1$ brane model. To infer its consistency with age of
our universe, we compare the age of old cosmological objects with
what computed using the best fit values for the model parameters.
When the universe is matter dominated, $w=0$, at $68\%$ level of
confidence, the combination of Gold sample SNIa, Cosmic Microwave
Background (CMB) shift parameter and SDSS databases provides
$\Omega_m=0.31_{-0.02}^{+0.02}$,
$\Omega_{\cal{C}}=0.05_{-0.01}^{+0.01}$,
$w_r=-1.40_{-0.20}^{+0.20}$, hence a spatially open universe with
$\Omega_k=0.21_{-0.08}^{+0.08}$. The same combination with SNLS
supernova observation gives $\Omega_m=0.28_{-0.02}^{+0.03}$,
$\Omega_{\cal{C}}=0.037_{-0.004}^{+0.003}$,
$w_r=-2.05_{-0.15}^{+0.15}$ consequently provides a spatially open
universe $\Omega_k=0.11_{-0.07}^{+0.10}$. These results obviously
seem to be in contradiction with the most recent WMAP results
indicating a flat
universe. \\
PACS numbers: 98.80.Es, 95.36.+x\\
\end{abstract}
\newpage
]
\section{Introduction}
Recent observations of type Ia supernovas (SNIa) suggest that the
expansion of the universe is accelerating
\cite{ris,permul,R04,Tonry}. As is well known all usual types of
matter with positive pressure generate attractive forces, which
decelerate the expansion of the universe. Given this, a dark energy
component with negative pressure was suggested to account for the
invisible fuel that drives the current acceleration of the universe.
There are a huge number of candidates for the dark energy component
in the literature (see, e.g.,
\cite{sa00,wein,lim04,cop06,arm00,pad03} for recent reviews), such
as a cosmological constant $\Lambda$ \cite{wein,bennett,peri,spe03},
an evolving scalar field (referred to by some as quintessence:
\cite{sa1,sa2,wet88,amen,peb88,cal03,arb05,wan00,per99,pag03,dor01,dor02,dor04},
the phantom energy, in which the sum of the pressure and energy
density is negative \cite{cal03,cal02,da03}, the quintom model
\cite{amen,zong}, the holographic dark energy \cite{li04}, the
Chaplygin gas \cite{be02,kam01}, and the Cardassion model
\cite{zong,bar05,da04}. Another approach dealing with this problem
is using the modified gravity by changing the Einstein-Hilbert
action. Some of models as $1/R$ and logarithmic models provide an
acceleration for the universe at the present time
\cite{bar05,rahvar07,sar}.
\\
In addition to the phenomenological modifications of the action, the
brane cosmology also implies modification on the dynamics of the
universe. Many different Brane models have been presented in the
recent years. In one of the best studied models, a codimension 1
thin brane with an infinite extra dimension has been investigated.
This brane is embedded in a bulk where its curvature to be negative
and its volume is finite \cite{RSII}. In this scenario, gravitation
is localized on a brane reproducing effectively four-dimensional
gravity at large distances due to the warp geometry of the
spacetime. Usually the brane is modeled as a distributional source
in the energy-momentum tensor (EMT) of zero thickness, and in this
case the cosmology has been obtained and analyzed in detail
\cite{Binetruy:1999ut,Ida:1999ui}. Recognizing the difficulty of
handling thick walls within relativity, already early authors
considered the idealization of a singular hypersurface as a thin
wall and tried to formulate its dynamics within general relativity
\cite{SLD}.
The new era of intense interests in thin shells and walls began
with the development of ideas related to phase transitions in
early universe and the formation of topological defects. Again,
mainly because of technical difficulties, strings and domain walls
were
assumed to be infinitesimally thin \cite{ViCv}.\\
Thereafter, interest in thin walls, or hypersurfaces of
discontinuity, received an impetus from the cosmology of early
universe. The formulation of dynamics of such singular
hypersurfaces was summed up in the modern terminology by Israel
\cite{israel}. Within the Sen-Lanczos-Israel (SLI) formalism, thin
shells are regarded as idealized zero thickness objects, with a
$\delta$-function singularity in their energy-momentum
and Einstein tensors.\\
In contrast to thin walls, thickness brings in new subtleties,
depending on how the thickness is defined and handled. Early
attempts to formulate thickness, being mainly motivated by the
outcome of the idea of late phase transition in cosmology
\cite{hill}, were concentrated on domain walls. Bonjour et al.,
studied a thick gravitating domain wall for a general potential
\cite{bon1,bon2}. Using general analytical arguments they have shown
 that all nontrivial solutions fall into two categories: those
 interpretable as an isolated domain wall with a cosmological event
 horizon, and those which are pure false vacuum de Sitter solutions.
  Also they have analyzed the domain wall in the presence of a cosmological
 constant finding the two kinds of solutions, wall and de Sitter,
 even in the presence of a negative cosmological constant.
 Silveria
\cite{Sil} studied the dynamics of a spherical thick domain wall
by appropriately defining an average radius for the wall. Widrow
\cite{Wid} used the Einstein-scalar equations for a static thick
domain wall with planar symmetry. He then took the zero-thickness
limit of his solution and showed that the orthogonal components of
the energy-momentum tensor  would vanish in that limit. Garfinkle
and Gregory \cite{GG} presented a modification of the Israel thin
shell equations by using an expansion of the coupled
Einstein-scalar field equations describing the thick gravitating
wall in powers of the thickness of the domain wall around the
well-known solution of the hyperbolic tangent kink for a
$\lambda\phi^{4}$ wall and concluded that the effect of thickness
at first approximation was effectively to reduce the energy
density of the wall compared to the thin case, leading to a faster
collapse of a spherical wall in vacuum. Others \cite{bar} applied
the expansion in the wall action and integrate it out
perpendicular to the wall to show that the effective action for a
thick domain wall in vacuum apart, from the usual Nambu term,
consists of a contribution proportional
to the induced Ricci curvature scalar.\\
Study of thick branes in the string inspired context of cosmology
began almost simultaneously with the study of thin branes, using
different approaches. Although in brane cosmology the interest is
in local behavior of gravity and the brane, most of the authors
take a planar brane for granted \cite{Ell}. However, irrespective
of the spacetime dimension and the motivation of having a wall or
brane, as far as the geometry of the problem is concerned, most of
the papers are based on a regular solution of Einstein equations
on a manifold with specified asymptotic behavior representing a
localized scalar field \cite{Whole}. Some authors use a smoothing
or smearing mechanism to modify the Randall-Sundrum ansatz
\cite{Csak,GhKo}. Authors in \cite{GhKo} introduce a thickness to
the brane by smoothing out the warp factor of a thin brane world
to investigate the stability of a thick brane. In another approach
to derive generalized Friedmann equations, the four-dimensional
effective brane quantities are obtained by integrating the
corresponding five-dimensional ones along the extra-dimension over
the brane thickness \cite {Moun}. These cosmological equations
describing a brane of finite thickness interpolate between the
case of an infinitely thick brane corresponding to the familiar
Kaluza-Klein picture and the opposite limit of an infinitely thin
brane giving the unconventional Friedmann equation, where the
energy density enters quadratically. The latter case is then made
compatible with the conventional cosmology at late times by
introducing and fine tuning a negative cosmological constant in
the bulk and an intrinsic positive tension in the brane
\cite{whole1}. A completely different approach based on the gluing
of a thick wall considered as a regular manifold to two different
manifolds on both sides of it was first suggested in
\cite{khak02}. The idea behind this suggestion is to understand
the dynamics of a localized matter distribution of any kind
confined within two metrically different spacetimes or matter
phases. Such a matching of three different manifolds is envisaged
to have many diverse applications in astrophysics, early universe,
and string cosmology. Authors in \cite{ghkhm} studied such a thick
brane to obtain the cosmological evolution of the brane.
 Navarro and Santiago \cite{Nav05} considered a
thick codimension 1 brane including a matter pressure component
along the extra dimension in the energy-momentum of the brane. By
integrating the 5D Einstein equations along the fifth dimension,
while neglecting the parallel derivatives of the metric in
comparison with the transverse ones, they write the equations
relating the values of the first derivatives of the metric at the
brane boundary with the integrated components of the brane
energy-momentum tensor. These, so called matching conditions are
then used to obtain the cosmological evolution of the brane  which
is of a non-standard type, leading to an accelerating universe for
special values of the model parameters. They show that when one
drops the infinitesimally thin idealization in the modelling of
the brane, gets non-standard cosmology on the brane.\\

In Section II we make a review over the model first introduced by
Navarro and Santiago, the cosmology of a thick codimension $1$
brane model, its free parameters and modified Friedman equation
which governs the background dynamics of the universe. Then we
show how this model can produce an accelerating universe. In
Section III we put some constraints on the parameters of model by
the background evolution, such as new Gold sample and Legacy
Survey of Supernova Type Ia data \cite{R04}, the combination with
the position of the observed acoustic angular scale on CMB and the
baryonic oscillation length scale. In Section  IV we study the age
of universe in the thick brane model,  we also compare the age of
the universe in this model with the age of old cosmological
structures in this section. Section V contains conclusions and
discussions of this work.

\section{Cosmology on the thick brane}
In this section we make a brief review over the model first
introduced by  Navarro and Santiago to study a thick brane. They
consider a thick brane in a Randall-Sundrum model with $Z_2$
symmetry. To get the cosmological behavior of the Braneworld, the 5D
metric is considered to be as follows: \be ds^2 = n^2(r,t)dt^2 -
a^2(r,t)d{\bf x}^2 - e^{2\phi(r,t)}dr^2 \ee where $r$ is the extra
dimension and the thick brane exists in $|r|< \epsilon$, so
$\epsilon$ is somehow the thickness of the brane. By the $Z_2$
symmetry we know that the core of the brane is placed at $r=0$. Now
we should write down the 5D Einstein equations in the bulk and using
the junction conditions derive the induced dynamics on the brane.
But first we need to know the energy momentum tensor of the brane
and the bulk. We  assume that in the bulk ($|r|>\epsilon$) the EMT
is just that of a cosmological constant,
$T^M_{N}=\delta^M_N\Lambda$. And for inside the brane,  a thick
brane means  we have let matter goes through the extra dimension.
So, matter is distributed within the thickness of the brane and the
5D energy momentum tensor can be written as follows: \be T^M_{N
\:(brane)}=diag(T^0_0,T^x_x,T^y_y,T^z_z,T^r_r)\ee where considering
the matter on the brane to be a perfect fluid, $T^0_0$ is the energy
density of matter, $T^x_x,T^y_y,T^z_z$ are the pressure of the
matter along the normal coordinates and
$T^r_r$ is the pressure along the extra dimension. \\
But, there is a problem to study this thick brane: what is the
suitable junction condition? In the thin wall approximation we
consider the brane to be a singular hypersurface with
$\delta$-function singularity in their energy-momentum and Einstein
tensors. From the Sen-Lanczos-Israel (SLI) formalism we know that
the difference of the 4D induced metric's derivatives, on both sides
of the thin brane is given by the energy momentum tensor of the
matter on the singular hypersurface. But in a thick brane model you
can not follow the same procedure, as the matter is distributed
along the extra dimension. So, what Navarro and Santiago have done
is to integrate over Einstein equations in the $|r|< \epsilon$
region to obtain the suitable matching conditions (they have
neglected the first derivative of the metric along the parallel
brane coordinates): \bea e^{-\phi}\frac{2 n^\prime}{n}|_\epsilon& =&
\frac{1}{M_\ast^3} \left[\frac{2}{3}\rho + p + \frac{1}{3}p_r
\right]\label{matching1}
\\
e^{-\phi}\frac{2 a^\prime}{a}|_\epsilon &=& \frac{1}{M_\ast^3}
\left[\frac{1}{3}\left(p_r-\rho\right)\right] \label{matching2} \eea
where prime denotes differentiation with respect to $r$,
${M_\ast^3}$ is the 5D fundamental mass and $\rho$, $p$ and $p_r$
respectively are the 4D energy density, longitudinal (along the
normal coordinates) and transverse 4D pressure (along the extra
dimension), derived by integrating over the EMT along the thickness
of the brane:

\bea \rho &\equiv&
\frac{1}{na^3|_{\epsilon}}\int_{-\epsilon}^\epsilon T_0^{0} n a^3
e^\phi dr \\
p&\equiv& -\frac{1}{na^3|_{\epsilon}}\int_{-\epsilon}^\epsilon
T_x^{x} n a^3 e^\phi dr
\\
p_r&\equiv& -\frac{1}{na^3|_{\epsilon}} \int_{-\epsilon}^\epsilon
T_r^{r} n a^3 e^\phi dr \eea Splitting  the brane EMT to the
constant ($\lambda$) and a time dependant part (with an arbitrary
but assumed constant equation of motion), we have:

\be \rho = \lambda + \rho_m \quad p = -\lambda + w \rho_m\quad p_r =
-{\lambda}_r +w_r \rho_m \ee

Now having the integrated EMT and the junction conditions we can
solve the 5D Einstein equations to obtain the cosmological
dynamics of the thick braneworld. Evaluating the $rr$ and $tr$
components of the 5D Einstein equations and keeping the terms up
to the first order of $\rho_m$  we get:

\bea &&3\left(
\frac{\ddot{a}}{a}+\frac{\dot{a}^2}{a^2}+\frac{k}{a^2}\right)
=\nonumber\\
&& \frac{1}{12 M_\ast^6} \Big[
2(\lambda+\lambda_r)^2+(\lambda+\lambda_r)(1-3w-4w_r)\rho_m\Big]
+\frac{\Lambda}{M_\ast^3} \label{friedtr}\nonumber\\ \eea \be
\dot{\rho}_m + 3\frac{\dot{a}}{a}\left(1 + w \right)\rho_m - w_r
\dot{\rho}_m=0\label{EMTcons} \ee where $k$ denotes the curvature
and can be $k=(0,1,-1)$ which respectively gives the spatially flat,
closed and open universes, $w$ and $w_r$ are the state parameters
related to the longitudinal and the transverse pressures. It can be
easily checked that if $w_r=0$ we obtain the standard cosmology with
a cosmological constant given by:\newline
$\Lambda_{eff}=\frac{1}{6M_\ast^6} \left(\lambda +
\lambda_r\right)^2 +\frac{\Lambda}{M_\ast^3}$) up to ${\cal
O}(\rho_m^2)$. But for a nonzero value of $w_r$ the cosmological
evolution on the 4D brane will be completely different. As it can be
seen in the  equation (\ref{EMTcons}), existing a non zero pressure
along the extra dimension violates the EMT conservation in the 4
dimensions: \be \dot{\rho}_m + 3\frac{\dot{a}}{a}\left(1 + w
\right)\rho_m = w_r \dot{\rho}_m\neq 0\label{EMTcons2} \ee One can
easily integrate equation (\ref{EMTcons2}) to get:
  \be \rho_m =
\rho_{m0}a^{-3\frac{1+w}{1-w_r}} \label{rho}\ee where $\rho_{m0}$ is
the present matter density. Inserting this result in equation
(\ref{friedtr}) and integrating the equation, assuming
$\Lambda_{eff}=0$, they obtain their generalized Friedman equation
as follows: \be \label{modFrid}H^2={\cal
C}a^{-4}+\frac{\lambda+\lambda_r}{18M_{\ast}^6}\left(1-w_r\right)
\rho_{m0}a^{-3\frac{1+w}{1-w_r}}-\frac k {a^2} \label{H2} \ee where
$H\equiv\frac {\dot{a}}{a}$ and $\cal{C}$ is an integration constant
coming from the extra dimension assumption. This equation shows that
having a pressure component of EMT along the extra dimension is
similar to having a matter with an effective parameter of state
defined as follows:
$$w_{eff}=\frac{w + w_r}{1-w_r}$$
Also it can be checked from equation (\ref{friedtr}) that even when
$\Lambda_{eff}=0$ we can have an accelerating universe
($w_{eff}<-1/3$), which means that in the case of having matter as
the only source of the Einstein equations ($w=0)$, the evolution of
the universe is an accelerating one. So the only required condition
is $w_r<-1/2$. Navarro and Santiago discuss that a KK mechanism can
produces such a negative pressure by the KK modes. So to have an
accelerating universe we require the matter to be confined in the
thickness of the brane which causes a negative $w_r$. They have not
studied the dynamics of the universe at early times, whether it is
comparable with standard model of cosmology  or not. As it can be
checked easily, having $w_r<-1/2$, all the time during the evolution
of the universe, in the radiation dominated era ($w=1/3$), equation
(\ref{modFrid}) shows that the second term in the right hand side
behaves as $a^{b}$ where $b>-8/3$, so in the early universe where
$a$ is very small, the first term is the dominated term, which
corresponds to the radiation dominated ear $a^{-4}$. So this model
is compatible with the standard model of cosmology at the early
universe. But the main difference is at the late time, where the
second term in the equation (\ref{modFrid}) is big enough and the
effects of the brane thickness becomes important.\\

In order to compare the predictions of this model with the
observational tests, we rewrite equation (\ref{H2}) as a function
of dimensionless parameters:

\bea \label{hub}&& H^2(z;\Omega_m,\Omega_{\cal{C}},w,w_r)
=H_0^2[\Omega_{\cal{C}}(1+z)^4\nonumber\\
&&+\Omega_m(1-w_r)(1+z)^{3\frac{1+w}{1-w_r}}-(\Omega_{tot}-1)(1+z)^2]
\eea where $M^2_p=6\frac{M^6_{\ast}}{\lambda+\lambda_r}$,
$\Omega_{\cal{C}}\equiv\frac{\cal{C}}{H^2_0}$, $z$ is the redshift,
$\Omega_m$ is the present matter density (for simplicity we discard
its zero indice) and
$\Omega_{tot}=\Omega_m(1-w_r)+\Omega_{\cal{C}}=1+\Omega_k$.

\begin{figure}
\epsfxsize=9.5truecm\epsfbox{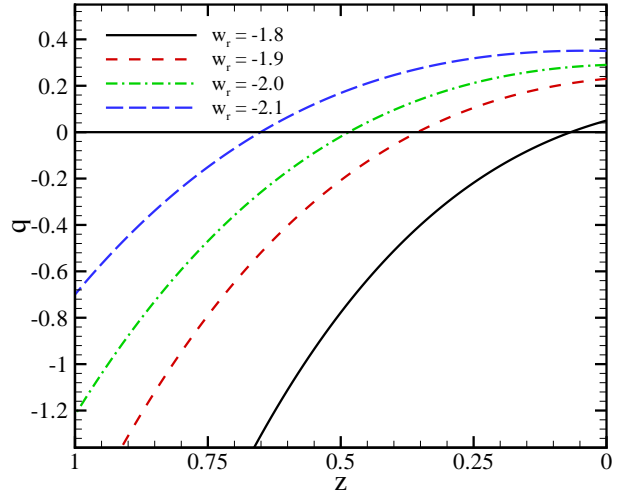} \narrowtext \caption{
Acceleration parameter ($q=\ddot{a}/aH_0^2$) in the thick brane
model as function of redshift for various values of $w_r$. Here
$\Omega_k=0.0$, $w=0.0$ and $\Omega_m=0.33$} \label{qbrane}
\end{figure}

In the case of having just energy density of matter as the only
source of the background dynamics of the universe, we have $w=0$.
 For an object with the redshift of $z$, using the null
geodesics in the FRW metric induced on the brane, comoving distance
is obtained as: \bea r(z;\Omega_m,\Omega_{\cal{C}},w_r,w) &=& {1
\over H_0\sqrt{|\Omega_k|}}\, \nonumber\\&&{\cal F} \left(
\sqrt{|\Omega_k|}\int_0^z\,
{dz' \over H(z')/H_0} \right) \label{comoving}\nonumber\\
 \eea
where
 \bea
 {\cal F}(x) &\equiv & (x,\sin x, \sinh
x)~\text{for}~k=(0,1,-1)\,
 \eea
and $H(z;\Omega_m,\Omega_{\cal{C}},w_r,w)$ is given by equation
(\ref{hub}). To see how this models gives an accelerating model we
study the behavior of the acceleration parameter
$(q=\ddot{a}/{H_0^2a})$. According to the equations (\ref{friedtr})
and (\ref{H2}) in this model accelerating parameter is derived as
follows: \be q=-\Omega_{\cal{C}}
a^{-4}-\frac{1}{2}(1+2w_r+3w)\Omega_ma^{-3(1+w_{eff})}\ee \label{q}
As it can be seen in figure (\ref{qbrane}), increasing $w_r$ causes
the universe to accelerate earlier.

Now an interesting question that arises is: "can this model predict
dynamics of universe?" or in another word, "what values of the model
parameter to be consistent with observational tests?"

In the forthcoming sections we will see what constraints to the
model described above are set by recent observations.

\section{Observational constraints on the model using background evolution of the universe}
In this section, at first we examine thick codimension $1$ brane
model by SNIa Gold sample and supernova Legacy Survey data. Then to
make the model parameter intervals more confined, we will combine
observational results of SNIa distance modules with power spectrum
of cosmic microwave background radiation and baryon acoustic
oscillation measured by Sloan Digital Sky survey. Table \ref{Tb1}
shows different priors on the model parameters used in the
likelihood analysis.
\begin{figure}[t]
\epsfxsize=9.0truecm
\begin{center}
\epsfbox{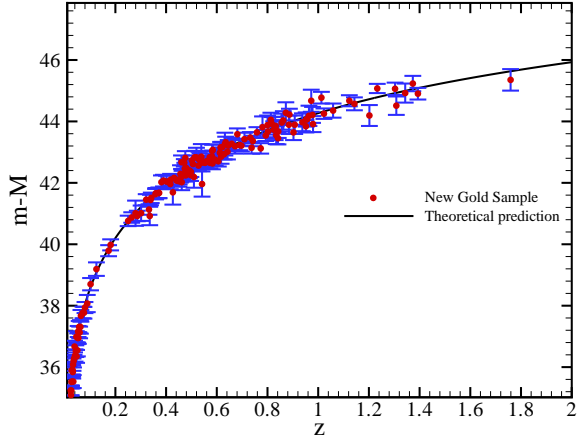} \narrowtext \caption{Distance modulus of the SNIa
new Gold sample in terms of redshift. Solid line shows the best fit
values with the corresponding parameters of $h=0.63$,
$\Omega_m=0.16^{+0.84}_{-0.03}$,
$\Omega_{\cal{C}}=0.20_{-0.19}^{+0.04}$
 and $w_r=-5.80^{+4.80}$ in $1 \sigma$ level of
confidence with $\chi^2_{min}/N_{d.o.f} =0.92$ for thick brane
model.} \label{modul1}
\end{center}
\end{figure}
\begin{table}
\begin{center}
\caption{\label{Tb1} Priors on the parameter space, used in the
likelihood analysis.}
\medskip
\begin{tabular}{|c|c|c|}
  {\rm Parameter}& Prior & \\  \hline
  $\Omega_{tot}=(1-w_r)\Omega_m+\Omega_{\cal{C}}$ & $1.00$ & {\rm Fixed}\\\hline
$\Omega_{tot}=(1-w_r)\Omega_m+\Omega_{\cal{C}}$ & - & {\rm
Free}\\\hline
 $\Omega_m$& $0.00-1.00$& {\rm Top hat}\\\hline
 $\Omega_bh^2$&$0.020\pm0.005$&{\rm Top hat (BBN)\cite{bbn}}\\\hline
 $h$&$-$&{\rm Free \cite{hst,zang}}\\\hline
$w_r$&$-100.00-0.00$&{\rm Top hat}\\\hline
 $w$&$0$&{\rm Fixed}\\
 \end{tabular}
\end{center}
\end{table}
 \subsection{Supernova Type Ia: Gold and SNLS Samples } \label{cobs}
The Supernova Type Ia experiments provided the main evidence of the
existence of dark energy. Since 1995 two teams of the {\it High-Z
Supernova Search} and the {\it Supernova Cosmology Project} have
discovered several type Ia supernovas at the high redshifts
\cite{per99,Schmidt}. Recently Riess et al.(2004) announced the
discovery of $16$ type Ia supernova with the Hubble Space Telescope.
They determined the luminosity distance of these supernovas and with
the previously reported algorithms, obtained a uniform $157$ Gold
sample of type Ia supernovas \cite{R04,Tonry,bar04}. Recently a new
data set of Gold sample with smaller systematic error containing
$156$ Supernova Ia has been released \cite{new}. In this work we use
this data set as new Gold sample SNIa.

More recently, the SNLS collaboration released the first year data
of its planned five-year Supernova Legacy Survey\cite{astier05}. An
important aspect to be emphasized on the SNLS data is that they seem
to be in a better agreement with WMAP results than the Gold sample
\cite{pad06}.

\begin{figure}[t]
\epsfxsize=9.0truecm
\begin{center}
\epsfbox{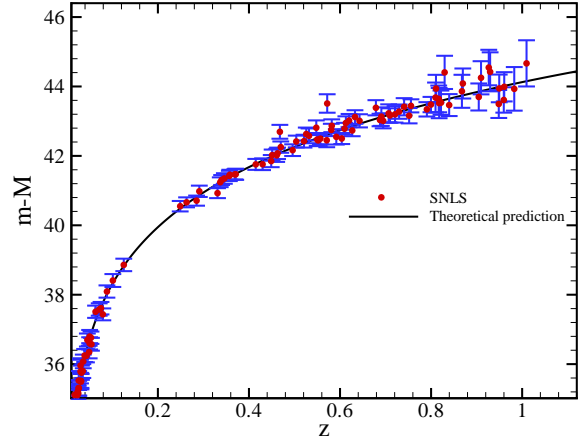} \narrowtext \caption{Distance modulus of the SNLS
supernova data in terms of redshift. Solid line shows the best fit
values with the corresponding parameters of $h=0.70$,
$\Omega_m=0.39^{+0.47}_{-0.26}$,
$\Omega_{\cal{C}}=0.054_{-0.053}^{+0.056}$
 and $w_r=-1.80^{+0.70}$  in $1 \sigma$ level of
confidence with $\chi^2_{min}/N_{d.o.f} =0.86$ for thick brane
model.} \label{modul2}
\end{center}
\end{figure}

We compare the predictions of the thick brane model for apparent
magnitude with new SNIa Gold sample and SNLS data set. The
observations of supernova measure essentially the apparent magnitude
$m$ including reddening, K correction, etc, which are related to the
(dimensionless) luminosity distance, $D_L$, of an object at redshift
$z$ through:
\begin{equation}
m={\mathcal{M}}+5\log{D_{L}(z;\Omega_m,\Omega_{\cal{C}},w_r,w)}
\label{m}
\end{equation} where
\begin{eqnarray}
\label{luminosity} D_L (z;\Omega_m,\Omega_{\cal{C}},w_r,w) &=&{(1+z)
\over \sqrt{|\Omega_k|}}\, {\cal F} \left(
\sqrt{|\Omega_k|}\int_0^z\, {dz'H_0\over H(z')} \right)
\end{eqnarray}
Also
\begin{eqnarray}
\label{m1}\mathcal{M} &=& M+5\log{\left(\frac{c/H_0}{1\quad
Mpc}\right)}+25
\end{eqnarray}
where $M$ is the absolute magnitude. The distance modulus, $\mu$, is
defined as:

\begin{eqnarray}
\mu\equiv
m-M&=&5\log{D_{L}(z;\Omega_m,\Omega_{\cal{C}},w_r,w)}\nonumber\\&&\quad
+5\log{\left(\frac{c/H_0}{1\quad Mpc}\right)}+25 \label{eq:mMr}
\end{eqnarray}
or

\begin{eqnarray}
\mu&=&5\log{D_{L}(z;\Omega_m,\Omega_{\cal{C}},w_r,w)}+\bar{M}
\end{eqnarray}

In order to compare the theoretical results with the observational
data, we must compute the distance modulus, as given by equation
(\ref{eq:mMr}). For this purpose, the first step is to compute the
quality of the fitting through the least squared fitting quantity
$\chi^2$ defined by:
\begin{eqnarray}\label{chi_sn}
&&\chi^2(\bar{M},\Omega_m,\Omega_{\cal{C}},w_r,w)=\nonumber\\
&&\sum_{i}\frac{[\mu_{obs}(z_i)-\mu_{th}(z_i;\Omega_m,\Omega_{\cal{C}},w_r,w,\bar{M})]^2}{\sigma_i^2}
\end{eqnarray}
where $\sigma_i$ is the observational uncertainty in the distance
modulus. To constrain the parameters of model, we use the Likelihood
statistical analysis:
\begin{eqnarray}
{\cal
L}(\bar{M},\Omega_m,\Omega_{\cal{C}},w_r,w)={\mathcal{N}}e^{-\chi^2(\bar{M},\Omega_m,\Omega_{\cal{C}},w_r,w)/2}
\end{eqnarray}
where ${\mathcal{N}}$ is a normalization factor. The parameter
$\bar{M}$ is a nuisance parameter and should be marginalized
(integrated out) leading to a new $\bar{\chi}^2$ defined as:
\begin{eqnarray}\label{mar2}
\bar{\chi}^2=-2\ln\int_{-\infty}^{+\infty}{\cal
L}(\bar{M},\Omega_m,\Omega_{\cal{C}},w_r,w)d\bar{M}
\end{eqnarray}

Using equations (\ref{chi_sn}) and (\ref{mar2}), we find:
\begin{eqnarray}\label{mar3}
\bar{\chi}^2(\Omega_m,\Omega_{\cal{C}},w_r,w)&=&\chi^2(\bar{M}=0,\Omega_m,\Omega_{\cal{C}},w_r,w)\nonumber\\&&
-\frac{B(\Omega_m,\Omega_{\cal{C}},w_r,w)^2}{C}+\ln(C/2\pi)
\end{eqnarray}
where
\begin{eqnarray}\label{mar4}
&&B(\Omega_m,\Omega_{\cal{C}},w_r,w)=\nonumber\\
&&\sum_{i}\frac{[\mu_{obs}(z_i)-\mu_{th}(z_i;\Omega_m,\Omega_{\cal{C}},w_r,w,\bar{M}=0)]}{\sigma_i^2}
\end{eqnarray}
and
\begin{eqnarray}\label{mar5}
C=\sum_{i}\frac{1}{\sigma_i^2}
\end{eqnarray}

Equivalent to marginalization is the minimization with respect to
$\bar{M}$. One can show that $\chi^2$ can be expanded in $\bar{M}$
as \cite{Nesseris04}:
\begin{eqnarray}\label{mar6}
\chi^2_{\rm
SNIa}(\Omega_m,\Omega_{\cal{C}},w_r,w)&=&\chi^2(\bar{M}=0,\Omega_m,\Omega_{\cal{C}},w_r,w)\nonumber\\&&
-2\bar{M}B+\bar{M}^2C
\end{eqnarray}
which has a minimum for $\bar{M}=B/C$:
\begin{eqnarray}\label{mar7}
\chi^2_{\rm SNIa}(\Omega_m,\Omega_{\cal{C}},w_r,w)&=&\chi^2(\bar{M}=0,\Omega_m,\Omega_{\cal{C}},w_r,w)\nonumber\\
&&-\frac{B(\Omega_m,\Omega_{\cal{C}},w_r,w)^2}{C}
\end{eqnarray}
Using equation (\ref{mar7}) we can find the best fit values of model
parameters as the values that minimize $\chi^2_{\rm
SNIa}(\Omega_m,\Omega_{\cal{C}},w_r,w)$. In the following analysis
we suppose matter domination era, $w=0$ unless stated otherwise. As
a simplest case we suppose a flat universe, $\Omega_k=0.0$. In this
situation, the best fit values for the parameters of the model are
$\Omega_m=0.01^{+0.12}_{-0.01}$ and $w_r=-10.00^{+4.69}$ with
$\chi^2_{min}/N_{d.o.f} =0.91$ at $1 \sigma$ level of confidence.
The corresponding value for the Hubble parameter at the minimized
$\chi^2$ is $h=0.63$ and since we have already marginalized over
this parameter we do not assign an error bar for it. The best fit
values for the parameters of model by using SNLS supernova data are
$\Omega_m=0.16^{+0.28}_{-0.16}$,
 and $w_r=-4.70^{+3.43}$ with $\chi^2_{min}/N_{d.o.f} =0.86$
at $1 \sigma$ level of confidence. The corresponding Hubble
parameter is $h=0.70$.

If we take account $\Omega_k$ as another free parameter, we would
have got $\Omega_{\cal{C}}$ as free parameter. The best fit values
for the parameters of model by using  supernova data are
$\Omega_m=0.16^{+0.84}_{-0.03}$,
$\Omega_{\cal{C}}=0.20_{-0.19}^{+0.04}$
 and $w_r=-5.80^{+4.80}$ with $\chi^2_{min}/N_{d.o.f} =0.92$
at $1 \sigma$ level of confidence. These values imply that
$\Omega_k=-0.28_{-1.62}^{+5.76}$. The best fit values for the
parameters of model by using SNLS supernova data are
$\Omega_m=0.39^{+0.47}_{-0.26}$,
$\Omega_{\cal{C}}=0.054_{-0.053}^{+0.056}$
 and $w_r=-1.80^{+0.70}$ with $\chi^2_{min}/N_{d.o.f} =0.87$
at $1 \sigma$ level of confidence. These values imply that
$\Omega_k=-0.14^{+1.34}$.

\begin{figure}[t]
\epsfxsize=9.5truecm\epsfbox{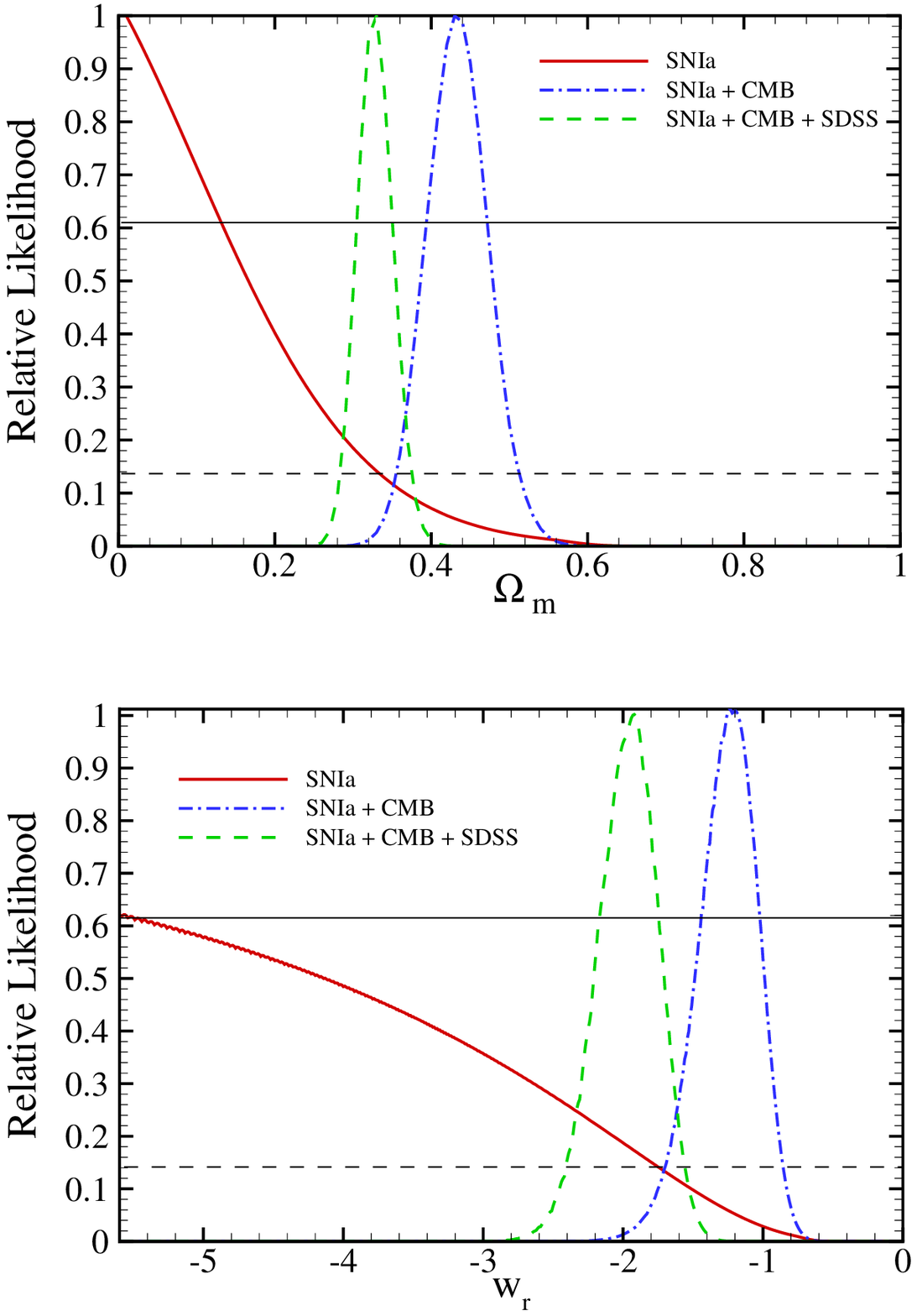} \narrowtext
\caption{Marginalized likelihood functions of two parameters of
model ($\Omega_m$ and $w_r$). The solid line corresponds to the
likelihood function of fitting the model with SNIa data (new Gold
sample), the dashdot line with the joint SNIa$+$CMB data and dashed
line corresponds to SNIa$+$CMB$+$SDSS. The intersections of the
curves with the horizontal solid and dashed lines give the bounds
with $1\sigma$ and $2\sigma$ level of confidence respectively. Here
$\Omega_{k}=0.0$ and $w=0.0$.} \label{like1}
 \end{figure}

\begin{figure}[t]
\epsfxsize=9.5truecm\epsfbox{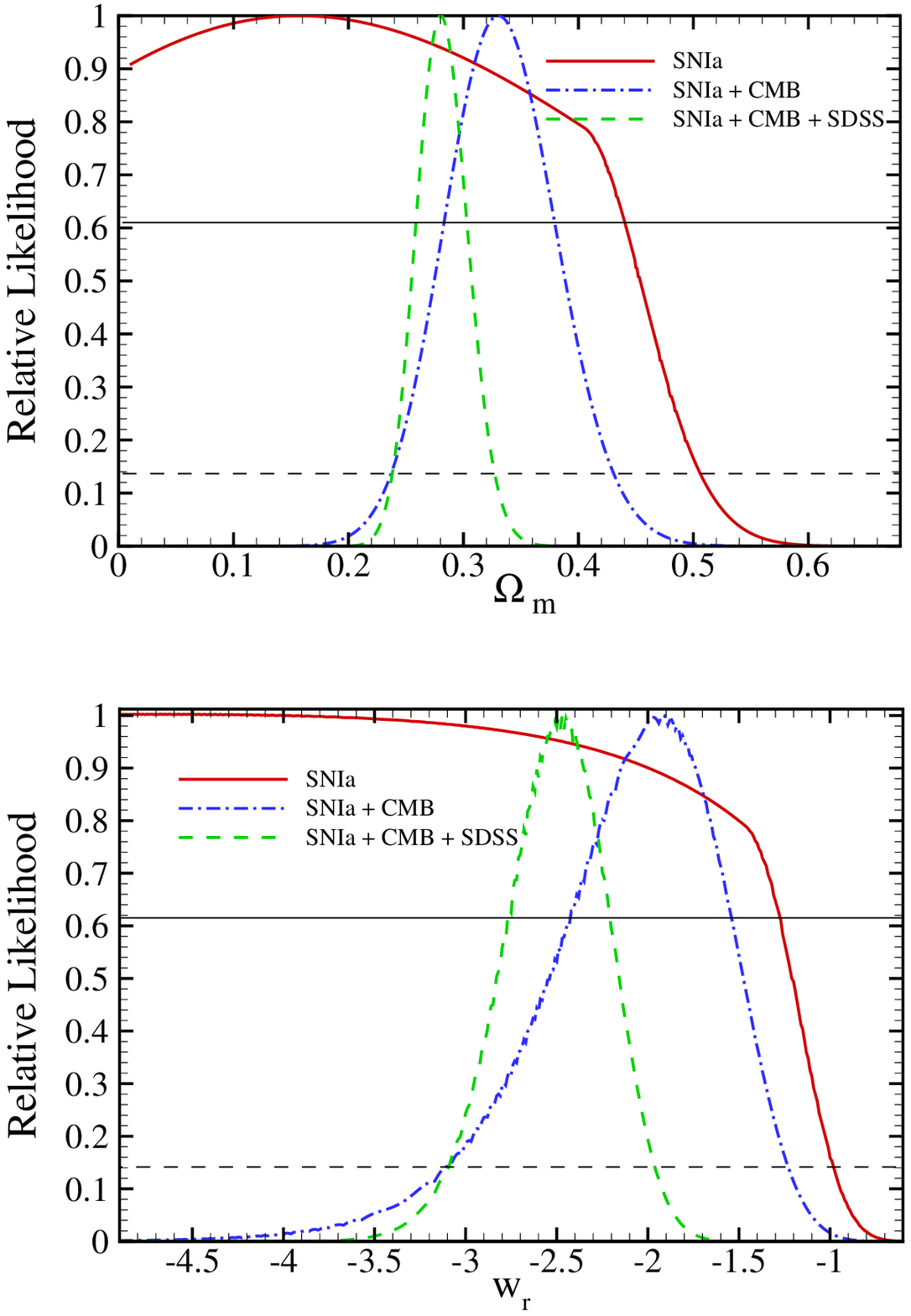} \narrowtext
\caption{Marginalized likelihood functions of two parameters of
thick brane model ($\Omega_m$ and $w_r$). The solid line corresponds
to the likelihood function of fitting the model with SNIa data
(SNLS), the dashdot line with the joint SNIa$+$CMB data and dashed
line corresponds to SNIa$+$CMB$+$SDSS. The intersections of the
curves with the horizontal solid and dashed lines give the bounds
with $1\sigma$ and $2\sigma$ level of confidence respectively. Here
$\Omega_{k}=0.0$ and $w=0.0$.} \label{like2}
 \end{figure}

\begin{figure}
\epsfxsize=9.0truecm\epsfbox{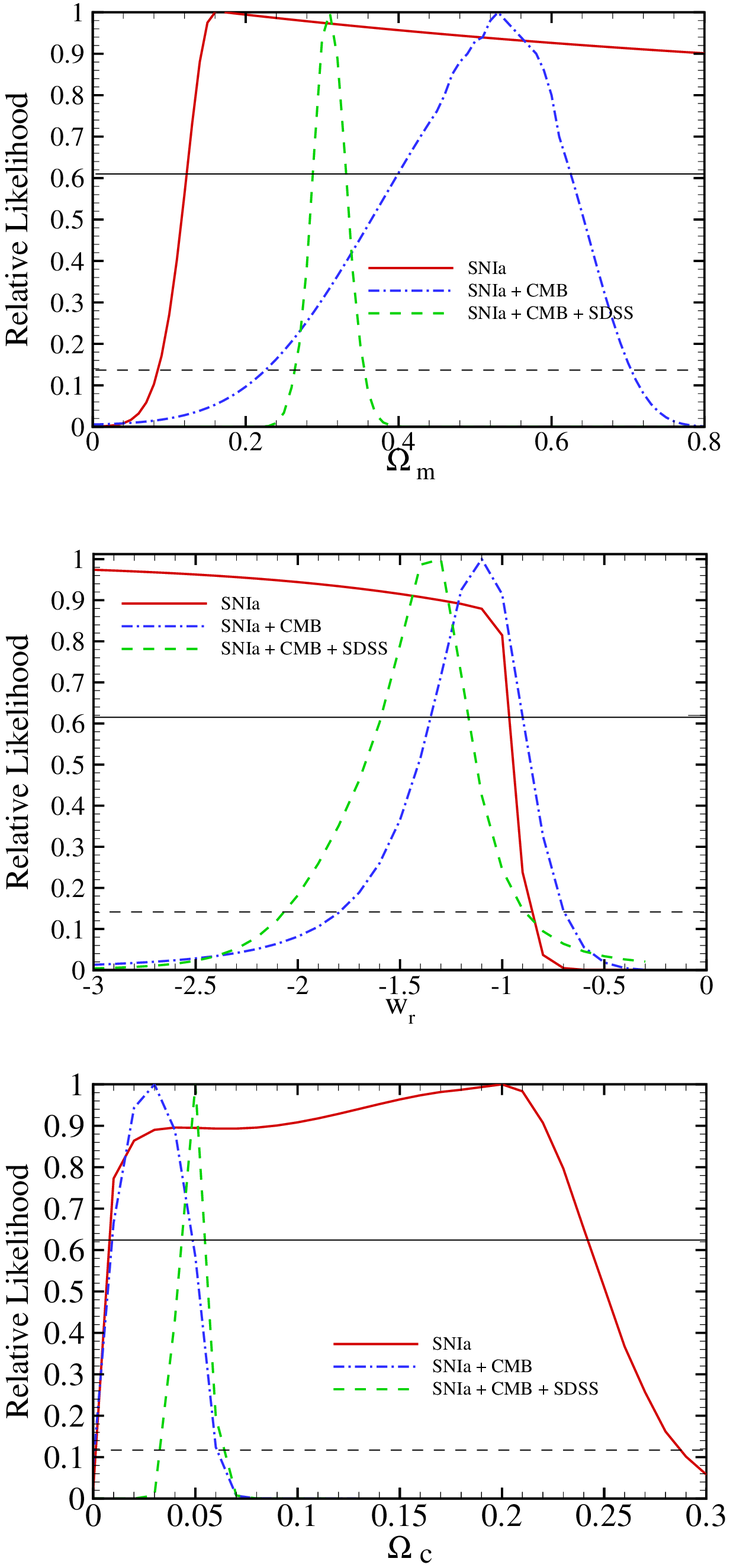} \narrowtext
\caption{Marginalized likelihood functions of three parameters of
model ($\Omega_m$, $w_r$ and $\Omega_{\cal{C}}$). The solid line
corresponds to the likelihood function of fitting the model with
SNIa data (new Gold sample), the dashdot line with the joint
SNIa$+$CMB data and dashed line corresponds to SNIa$+$CMB$+$SDSS.
The intersections of the curves with the horizontal solid and dashed
lines give the bounds with $1\sigma$ and $2\sigma$ level of
confidence respectively. Here we take $w=0.0$.} \label{like3}
 \end{figure}

\begin{figure}
\epsfxsize=9.0truecm\epsfbox{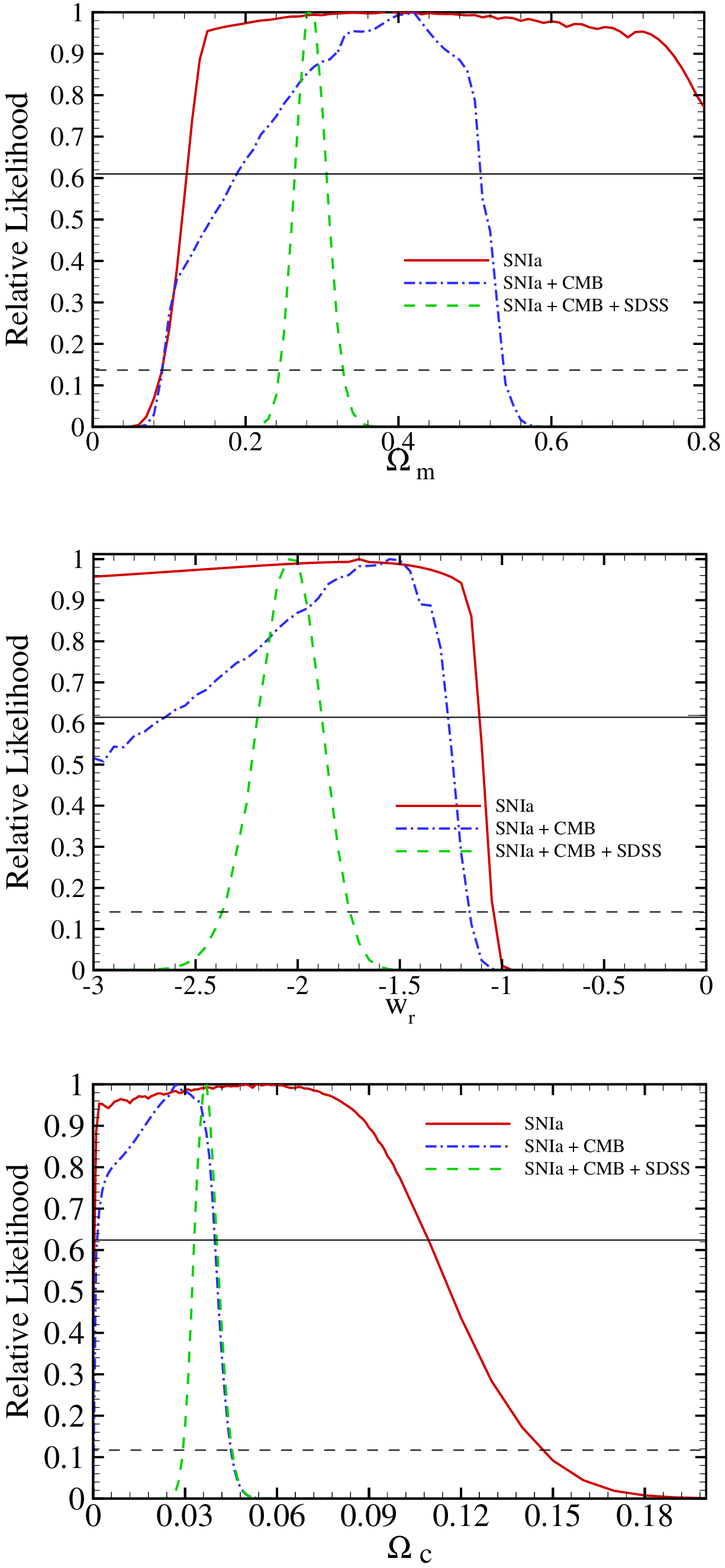} \narrowtext
\caption{Marginalized likelihood functions of three parameters of
model ($\Omega_m$, $w_r$ and $\Omega_{\cal{C}}$). The solid line
corresponds to the likelihood function of fitting the model with
SNIa data (SNLS), the dashdot line with the joint SNIa$+$CMB data
and dashed line corresponds to SNIa$+$CMB$+$SDSS. The intersections
of the curves with the horizontal solid and dashed lines give the
bounds with $1\sigma$ and $2\sigma$ level of confidence
respectively. Here we take $w=0.0$.} \label{like4}
 \end{figure}

 Figures (\ref{modul1}) and
(\ref{modul2}) show the comparison of the theoretical prediction of
distance modulus by using the best fit values of model parameters
and observational values from new Gold sample and SNLS supernova,
respectively. Figures (\ref{like1}), (\ref{like2}), (\ref{like3})
and (\ref{like4}) show relative likelihood for free parameters of
thick brane model.

\subsection{Combined analysis: SNIa$+$CMB$+$SDSS} \label{cmb}

Before last scattering, the photons and baryons are tightly coupled
by Compton scattering and behave as a fluid. The oscillations of
this fluid, occurring as a result of the balance between the
gravitational interactions and the photon pressure, lead to the
familiar spectrum of peaks and troughs in the averaged temperature
anisotropy spectrum which we measure today. The odd peaks correspond
to maximum compression of the fluid, the even ones to rarefaction
\cite{hu96}. In an idealized model of the fluid, there is an
analytic relation for the location of the $m$-th peak: $l_m \approx
ml_A$ \cite{Hu95,hu00} where $l_A$ is the acoustic scale which may
be calculated analytically and depends on both pre- and
post-recombination physics as well as the geometry of the universe.
The acoustic scale corresponds to the Jeans length of photon-baryon
structures at the last scattering surface some $\sim 379$ Kyr after
the Big Bang \cite{spe03}. The apparent angular size of acoustic
peak can be obtained by dividing the comoving size of sound horizon
at the decoupling epoch $r_s(z_{dec})$ by the comoving distance of
observer to the last scattering surface $r(z_{dec})$:
\begin{equation}
\theta_A =\frac{\pi}{l_A}\equiv {{r_s(z_{dec})}\over r(z_{dec}) }
\label{eq:theta_s}
\end{equation}
The size of sound horizon at the numerator of equation
(\ref{eq:theta_s}) corresponds to the distance that a
perturbation of pressure can travel from the beginning of
universe up to the last scattering surface and is given by:
 \bea
&&r_{s}(z_{dec};\Omega_m,\Omega_{\cal{C}},w_r,w) \nonumber\\
&&= {1 \over H_0\sqrt{|\Omega_k|}}\times {\cal F} \left(
\sqrt{|\Omega_k|}\int_{z_{dec}}^ {\infty} {v_s(z')dz' \over
H(z')/H_0} \right) \label{sh}
 \eea
where $v_s(z)^{-2}=3 + 9/4\times\rho_b(z)/\rho_{rad}(z)$ is the
sound velocity in the unit of speed of light from the big bang up to
the last scattering surface \cite{dor01,Hu95} and the redshift of
the last scattering surface, $z_{dec}$, is given by \cite{Hu95}:
\begin{eqnarray}\label{dec}
z_{dec} &=& 1048\left[ 1 + 0.00124(\omega_b)^{-0.738}\right]\left[
1+g_1(\omega_m)^{g_2}\right]\nonumber \\
g_1 &=& 0.0783(\omega_b)^{-0.238}\left[1+39.5(\omega_b)^{0.763}\right]^{-1}\nonumber \\
g_2 &=& 0.560\left[1 + 21.1(\omega_b)^{1.81}\right]^{-1}
\end{eqnarray}
where $\omega_m\equiv\Omega_mh^2$, $\rho_{rad}$ is the radiation
density and $\omega_b\equiv\Omega_bh^2$ ($\Omega_b$ is the present
baryonic density). Changing the parameters of the model can change
the size of apparent acoustic peak and subsequently the position of
$l_A\equiv \pi/\theta_A$ in the power spectrum of temperature
fluctuations on CMB. The simple relation $l_m\approx ml_A$ however
does not hold very well for the first peak although it is better for
higher peaks. Driving effects from the decay of the gravitational
potential as well as contributions from the Doppler shift of the
oscillating fluid introduce a shift in the spectrum. A good
parameterizations for the location of the peaks and troughs is given
by \cite{dor02,hu00}
\begin{equation}\label{phase shift}
l_m=l_A(m-\phi_m)
\end{equation}
where $\phi_m$ is phase shift determined predominantly by
pre-recombination physics, and are independent of the geometry of
the Universe. The location of  first acoustic peak can be determined
in model by equation (\ref{phase shift}) with
$\phi_1(\omega_m,\omega_b)\simeq 0.27$ \cite{dor02,hu00}. Instead of
the peak locations of power spectrum of CMB, one can use another
model independent parameter which is so-called shift parameter
${\cal R}$ as:
\begin{equation}
{\cal R}\propto\frac{l_1^{flat}}{l_1}
\end{equation}
where $l_1^{flat}$ corresponds to flat pure-CDM model with
$\Omega_m=1.0$ and the same $\omega_m$ and $\omega_b$ as the
original model.  It is easily shown that shift parameter is as
follows \cite{bond97}:
\begin{equation}\label{shift_th}
\label{shift} {\cal R}=
\sqrt{\Omega_m}\frac{D_L(z_{dec};\Omega_m,\Omega_{\cal{C}},w_r,w)}{(1+z_{dec})}
\end{equation}
The observational results of CMB experiments correspond to a shift
parameter of ${\cal R}=1.716\pm0.062$ (given by WMAP, CBI, ACBAR)
\cite{spe03,pearson03}. One of the advantages of using the parameter
${\cal R}$ is that it is independent of Hubble constant. In order to
put constraint on the model from CMB, we compare the observed shift
parameter with that of model using likelihood statistic as
\cite{bond97}:
\begin{equation}
{\cal{L}}\sim e^{-\chi_{\rm CMB}^2/2}
\end{equation}
where \begin{equation}\label{chi_cmb} \chi_{\rm
CMB}^2=\frac{\left[{\cal R}_{{\rm obs}}-{\cal R}_{{\rm
th}}\right]^2}{\sigma_{\rm CMB}^2}
\end{equation}
where ${\cal R}_{{\rm th}}$ and ${\cal R}_{{\rm obs}}$ are the
theoretical shift parameter, determined using equation
(\ref{shift_th}), and the observed one, respectively.

 The large scale correlation function measured from $46,748$
{\it Luminous Red Galaxies} (LRG) spectroscopic sample of the SDSS
(Sloan Digital Sky Survey) includes a clear peak at about  $100$ Mpc
$h^{-1}$ \cite{eisenstein05}. This peak was identified with the
expanding spherical wave of baryonic perturbations originating from
acoustic oscillations at recombination. The comoving scale of this
shell at recombination is about 150Mpc in radius. In other words
this peak has an excellent match to the predicted shape and the
location of the imprint of the recombination-epoch acoustic
oscillation on the low-redshift clustering matter
\cite{eisenstein05}. A dimensionless and independent of $H_0$
version of SDSS observational parameter is:
\begin{eqnarray} \label{lss1}
{\cal A} &=&D_V(z_{\rm sdss})\frac{\sqrt{\Omega_mH_0^2}}{z_{\rm
sdss}}\nonumber\\
&=&\sqrt{\Omega_m}\left[\frac{H_0D_L^2(z_{\rm
sdss};\Omega_m,\Omega_{\cal{C}},w_r,w)}{H(z_{\rm
sdss};\Omega_m,\Omega_{\cal{C}},w_r,w)z_{\rm sdss}^2(1+z_{\rm
sdss})^2}\right]^{1/3}\nonumber\\
\end{eqnarray}
where $D_V(z_{\rm sdss})$ is characteristic distance scale of the
survey with mean redshift $z_{\rm sdss}$
\cite{eisenstein05,blak03,ness06}. We use the robust constraint on
the thick brane model using the value of ${\cal A}=0.469\pm0.017$
from the LRG observation at $z_{\rm sdss} = 0.35$
\cite{eisenstein05}. This observation permits the addition of one
more term in the $\chi^2$ of equations (\ref{mar7}) and
(\ref{chi_cmb}) to be minimized with respect to $H(z)$ model
parameters. This term is:
\begin{equation}
\chi_{\rm SDSS}^2=\frac{\left[{\cal A}_{\rm obs}-{\cal A}_{\rm
th}\right]^2}{\sigma_{\rm SDSS}^2}
\end{equation}
This is the third observational constraint for our analysis.

\begin{figure}[t]
\epsfxsize=9.0truecm\epsfbox{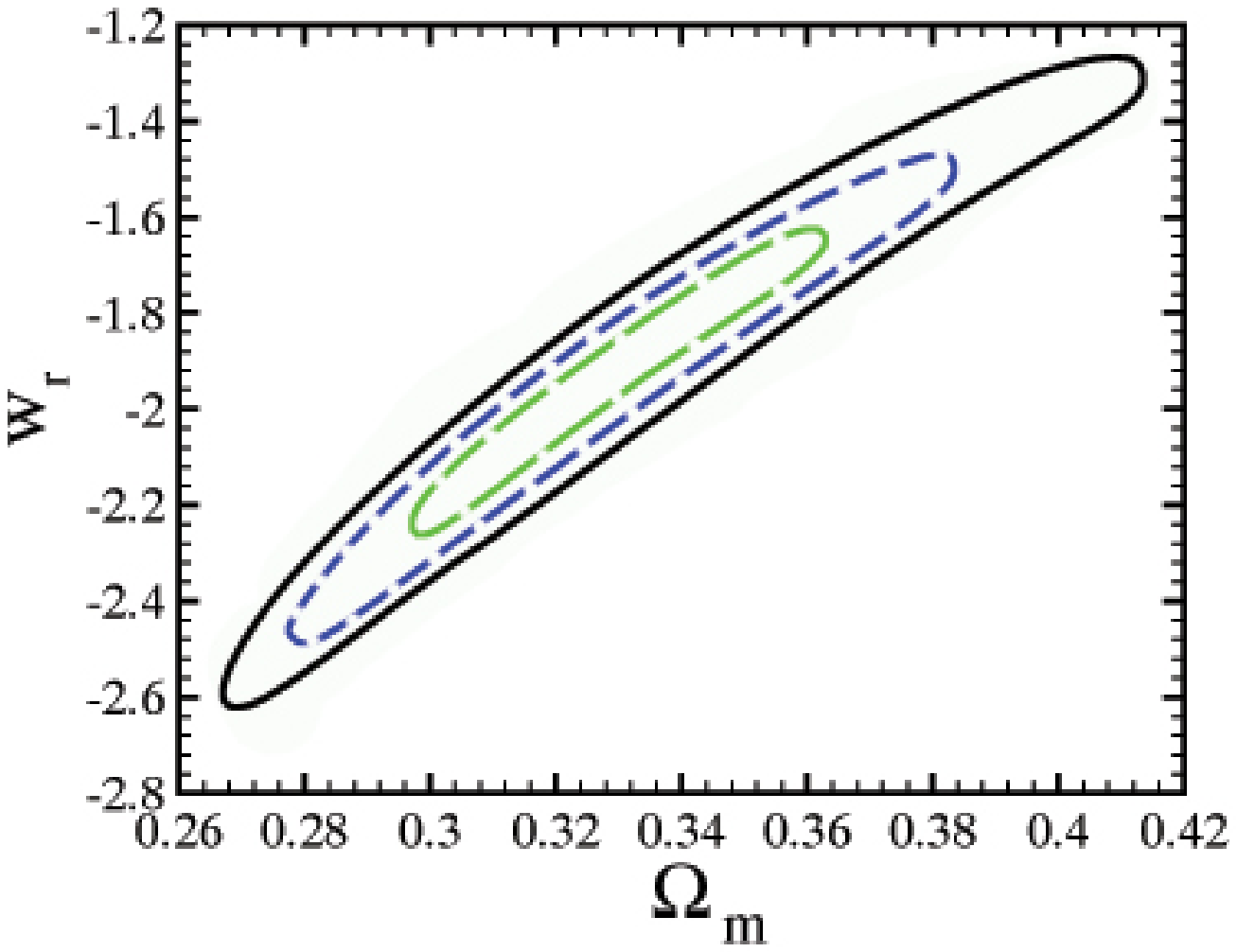} \narrowtext \caption{Joint
confidence intervals of $\Omega_m$ and $w_r$, fitted with SNIa new
Gold sample$+$CMB$+$SDSS. Solid line, dashed line and long dashed
line correspond to $3\sigma$, $2\sigma$ and $1\sigma$ level of
confidence, respectively. Here $\Omega_{k}=0.0$ and $w=0.0$.}
\label{jlike1}
 \end{figure}

\begin{figure}[t]
\epsfxsize=9.0truecm\epsfbox{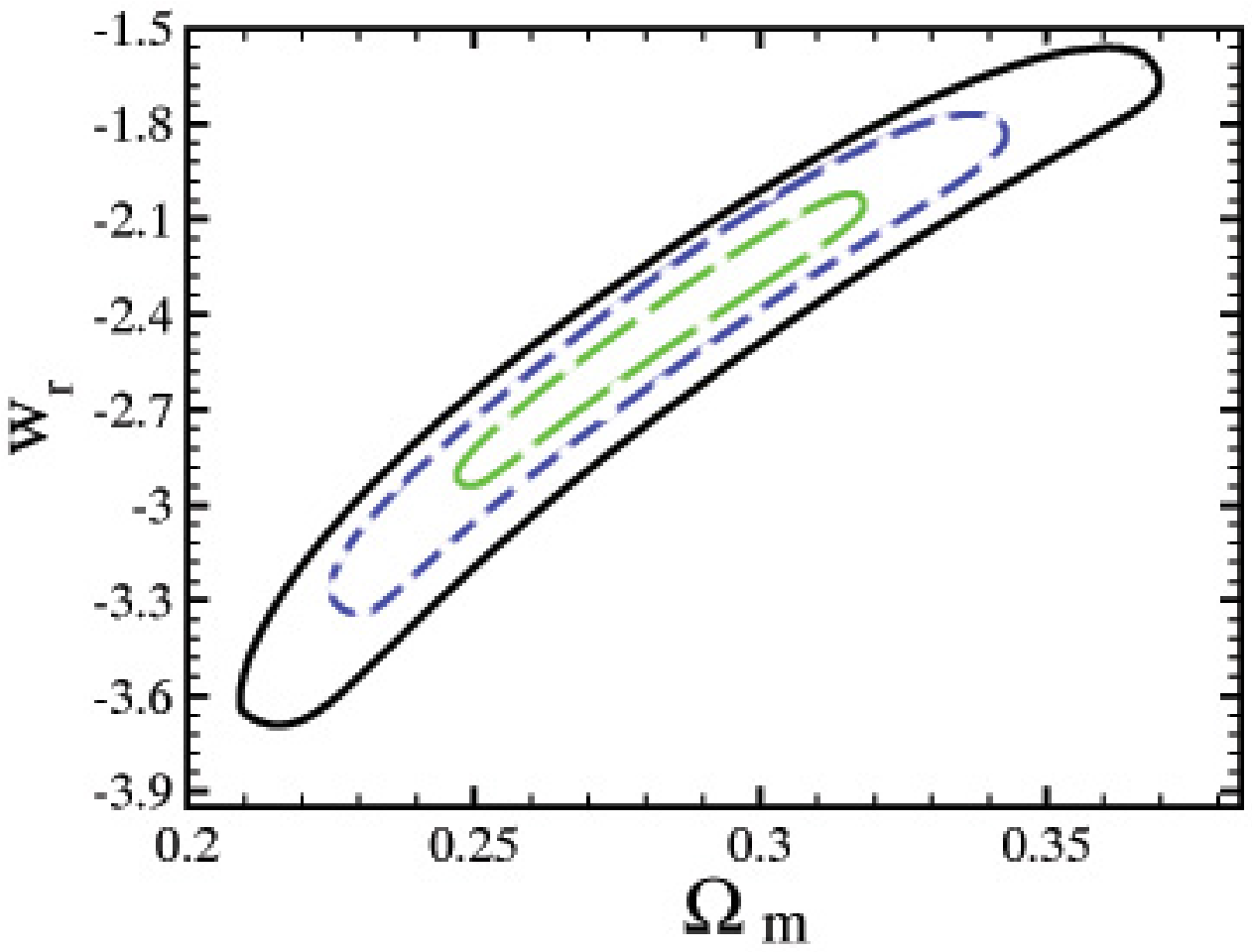} \narrowtext \caption{Joint
confidence intervals of $\Omega_m$ and $w_r$, fitted with SNIa
SNLS$+$CMB$+$SDSS. Solid line, dashed line and long dashed line
correspond to $3\sigma$, $2\sigma$ and $1\sigma$ level of
confidence, respectively. Here $\Omega_{k}=0.0$ and $w=0.0$.}
\label{jlike2}
 \end{figure}

\begin{figure}[t]
\epsfxsize=8.0truecm\epsfbox{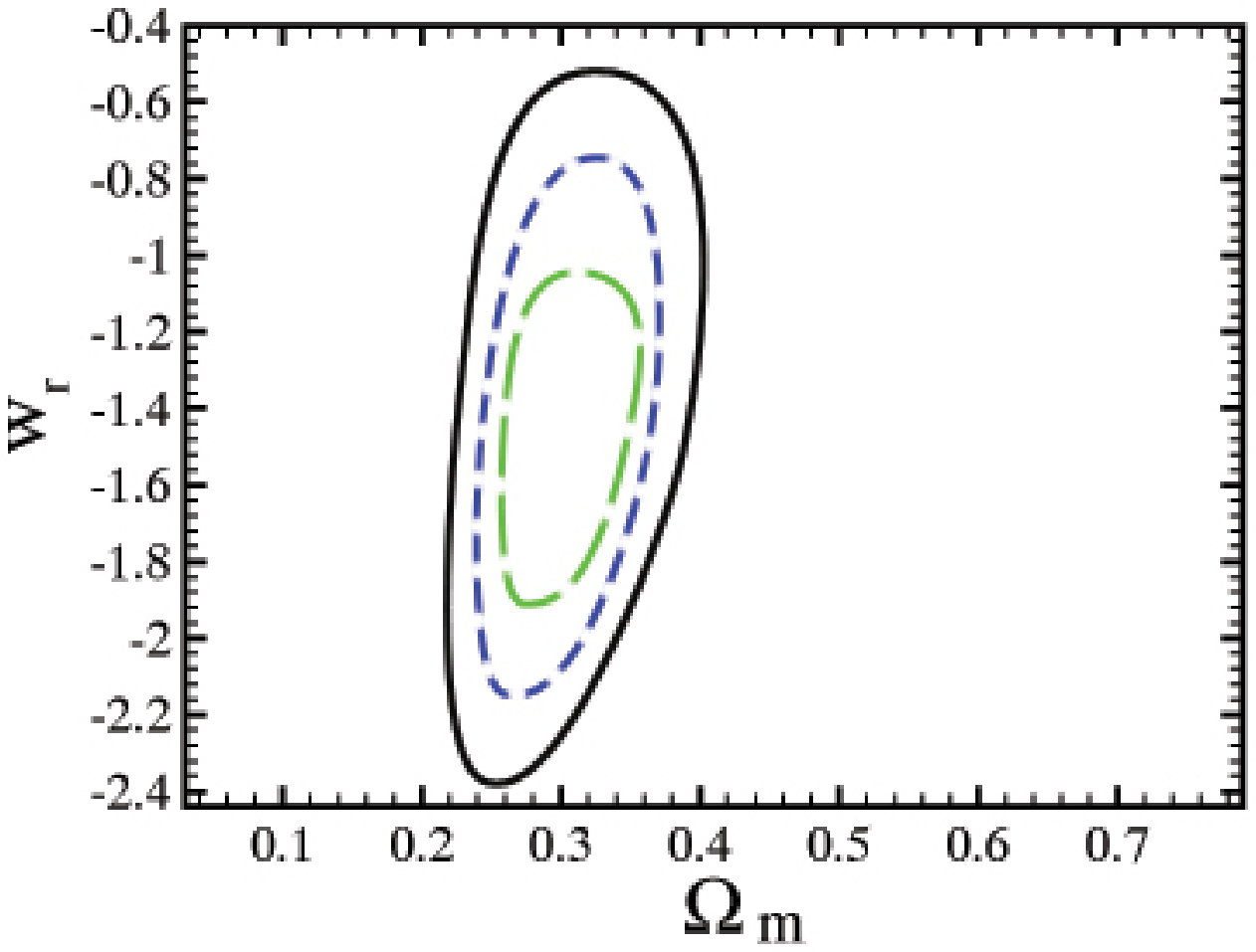} \narrowtext \caption{Joint
confidence intervals of $\Omega_m$ and $w_r$, fitted with SNIa new
Gold sample$+$CMB$+$SDSS. Solid line, dashed line and long dashed
line correspond to $3\sigma$, $2\sigma$ and $1\sigma$ level of
confidence, respectively. Here we supposed $w=0.0$.} \label{jlike3}
 \end{figure}

\begin{figure}[t]
\epsfxsize=8.0truecm\epsfbox{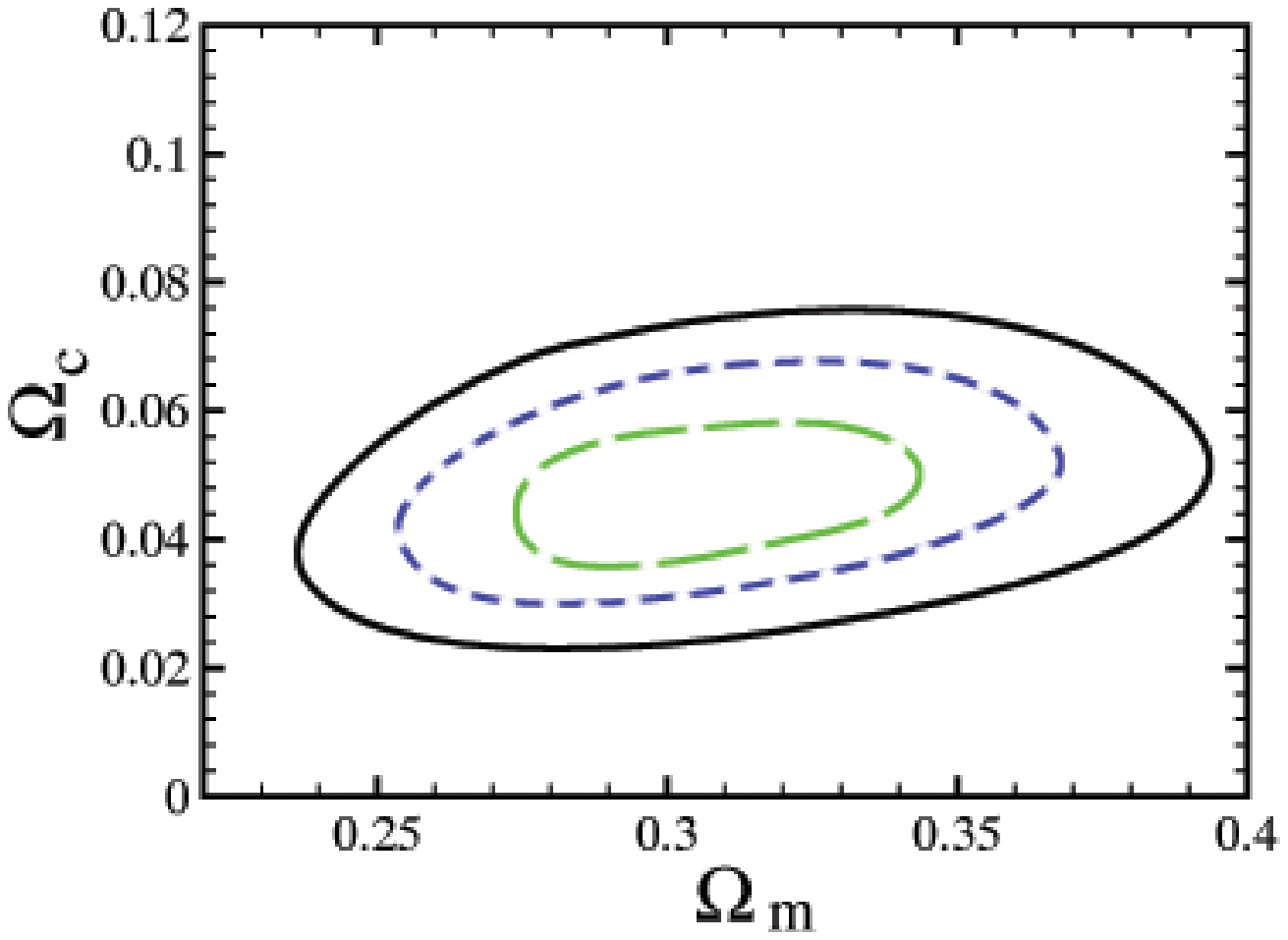} \narrowtext \caption{Joint
confidence intervals of $\Omega_m$ and $\Omega_{\cal{C}}$, fitted
with SNIa new Gold sample$+$CMB$+$SDSS. Solid line, dashed line and
long dashed line correspond to $3\sigma$, $2\sigma$ and $1\sigma$
level of confidence, respectively. Here we fixed $w=0.0$.}
\label{jlike4}
 \end{figure}

\begin{figure}[t]
\epsfxsize=8.0truecm\epsfbox{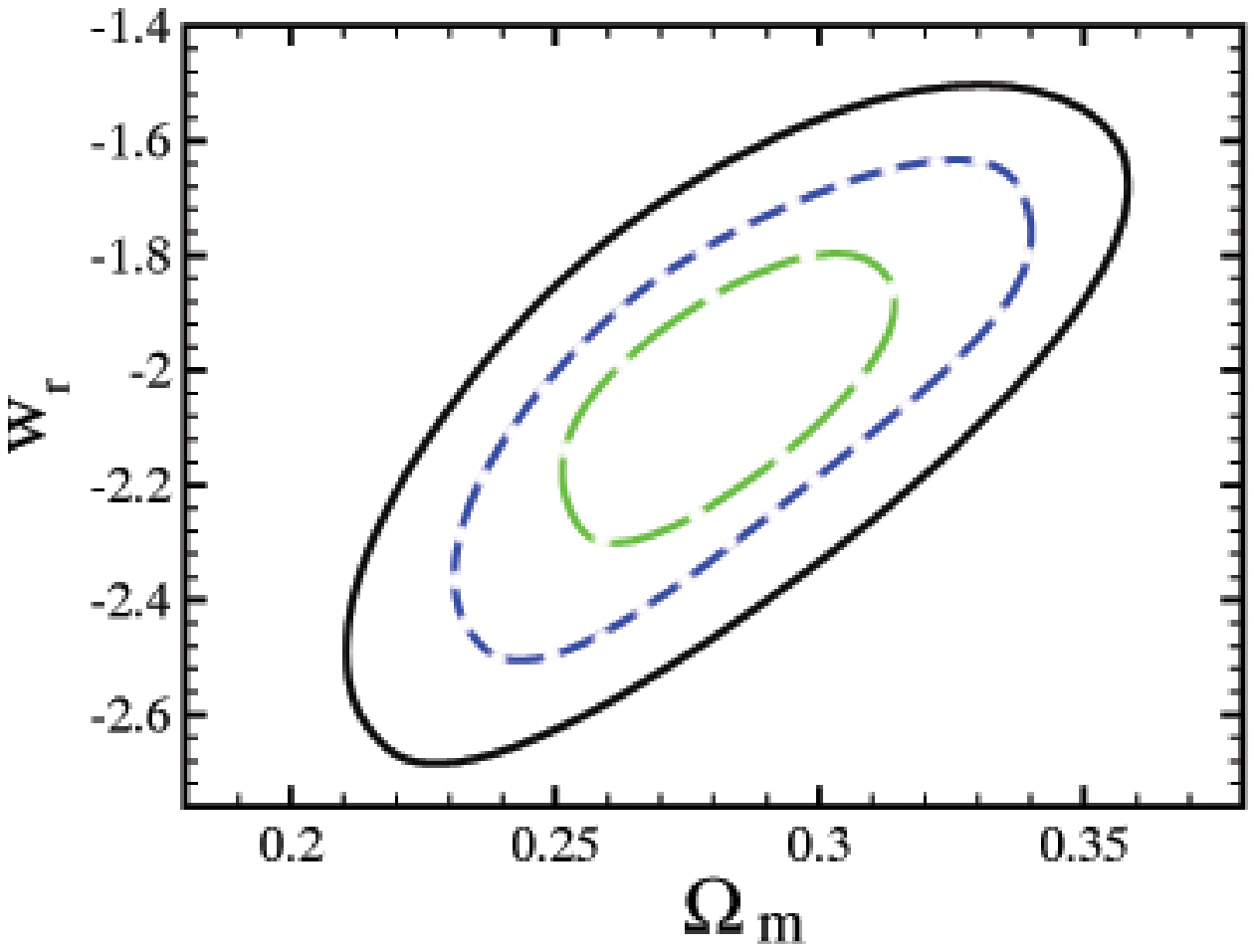} \narrowtext \caption{Joint
confidence intervals of $\Omega_m$ and $w_r$, fitted with SNIa
SNLS$+$CMB$+$SDSS. Solid line, dashed line and long dashed line
correspond to $3\sigma$, $2\sigma$ and $1\sigma$ level of
confidence, respectively. Here we fixed $w=0.0$.} \label{jlike5}
 \end{figure}

\begin{figure}[t]
\epsfxsize=8.0truecm\epsfbox{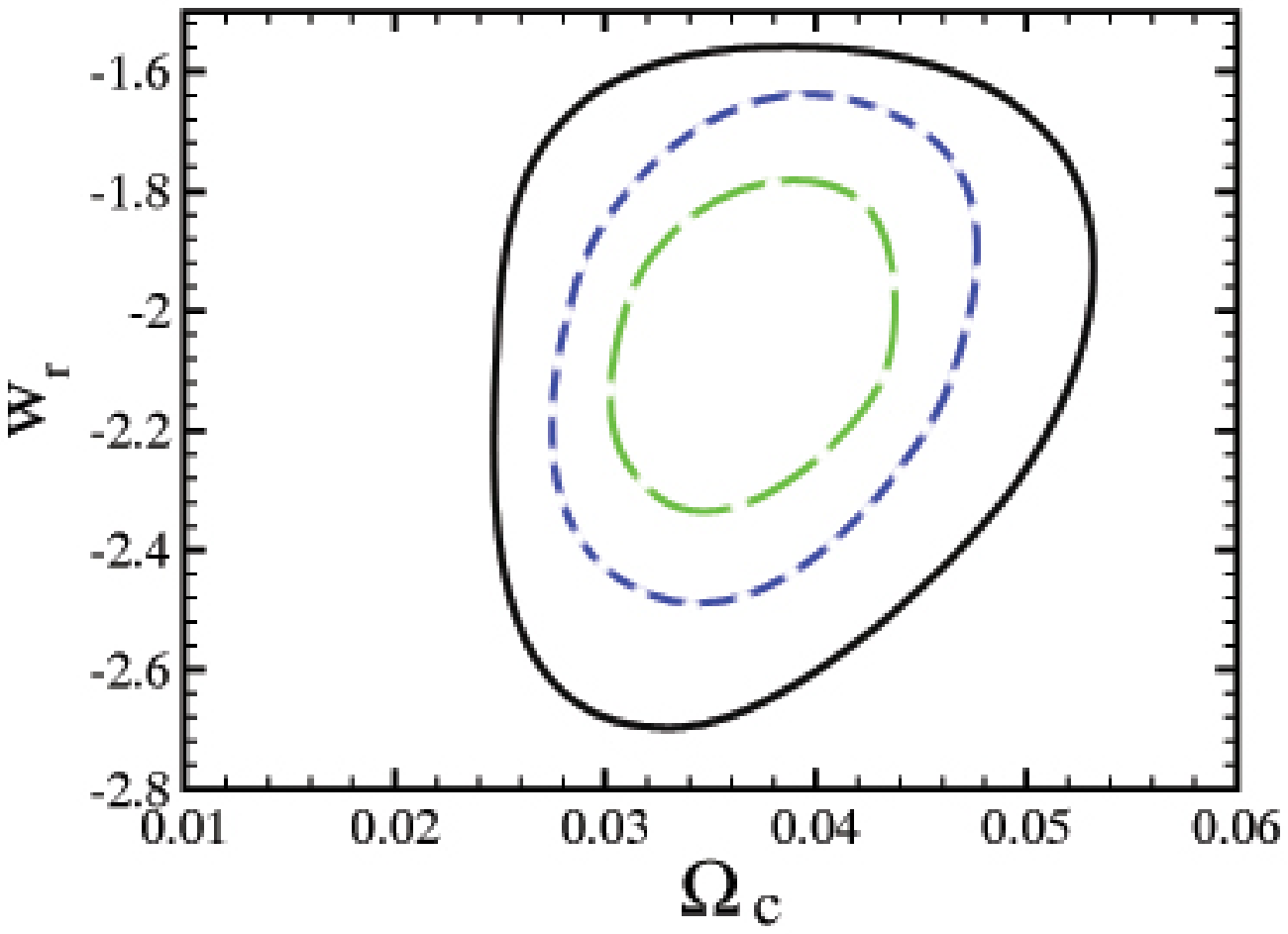} \narrowtext \caption{Joint
confidence intervals of  $\Omega_{\cal{C}}$ and $w_r$, fitted with
SNIa SNLS$+$CMB$+$SDSS. Solid line, dashed line and long dashed line
correspond to $3\sigma$, $2\sigma$ and $1\sigma$ level of
confidence, respectively. Here we fixed $w=0.0$.} \label{jlike6}
 \end{figure}

In what follows we perform a combined analysis of SNIa, CMB and SDSS
to constrain the parameters of the thick brane model by minimizing
the combined $\chi^2 = \chi^2_{\rm {SNIa}}+\chi^2_{{\rm
CMB}}+\chi^2_{{\rm SDSS}}$. The best values of the model parameters
from the fitting with the corresponding error bars from the
likelihood function marginalizing over the Hubble parameter in the
multidimensional parameter space in flat universe and domination
matter epoch are: $\Omega_m=0.33_{-0.03}^{+0.02}$ and
$w_r=-1.92^{+0.18}_{-0.25}$ at $1\sigma$ confidence level with
$\chi^2_{min}/N_{d.o.f}=1.02$. The Hubble parameter corresponding to
the minimum value of $\chi^2$ is $h=0.64$. Here we obtain an age of
$15.62_{-3.91}^{+3.14}$ Gyr for the universe (see section IV for
more details). Using the SNLS data, the best fit values of model
parameters are: $\Omega_m=0.28_{-0.02}^{+0.02}$ and
$w_r=-2.47^{+0.27}_{-0.30}$ at $1\sigma$ confidence level with
$\chi^2_{min}/N_{d.o.f}=0.86$. Age of universe calculating with the
best fit parameters is $15.23_{-4.07}^{+4.04}$ (see next section).
Table \ref{tab2} indicates the best fit values for the cosmological
parameters with one and two $\sigma$ level of confidence.
 In general case, the following values for the free parameters maximize
 likelihood probability: $\Omega_m=0.31_{-0.02}^{+0.02}$, $\Omega_{\cal{C}}=0.05_{-0.01}^{+0.01}$ and
$w_r=-1.40^{+0.20}_{-0.20}$ at $1\sigma$ confidence level stats
$\Omega_k=+0.21_{-0.08}^{+0.08}$. SNLS SNIa$+$CMB$+$SDSS give:
$\Omega_m=0.28_{-0.02}^{+0.03}$,
$\Omega_{\cal{C}}=0.037_{-0.004}^{+0.003}$ and
$w_r=-2.05^{+0.15}_{-0.15}$ at $1\sigma$ confidence level
demonstrate $\Omega_k=+0.11_{-0.07}^{+0.10}$. Tables \ref{tab3} and
\ref{tab4} give the best fit values for free parameters and age of
universe computing with these values. Joint confidence intervals in
free parameter spaces are shown in figures (\ref{jlike1}),
(\ref{jlike2}), (\ref{jlike3}), (\ref{jlike4}), (\ref{jlike5}) and
(\ref{jlike6}).

\begin{table}[t]
\begin{center}
\caption{\label{tab2} The best values for the parameters of a thick
brane model with the corresponding age for the universe from fitting
with SNIa from new Gold sample and SNLS data, SNIa+CMB and
SNIa+CMB+SDSS experiments at one and two $\sigma$ confidence level.
Here we suppose $\Omega_{k}=0.0$ and $w=0.0$. }
\begin{tabular}{|c|c|c|c|}
Observation & $\Omega_m$ & $w_r$ & Age (Gyr)
\\ \hline
  &&& \\
 & $0.01^{+0.12}_{-0.01}$&$-10.00^{+4.69}$ &  \\ 
 SNIa(new Gold)&&&$8.06 ^{+3.14}$\\
 & $0.01^{+0.33}_{-0.01}$&$-10.00^{+98.31}$&
\\ &&&\\ \hline
&&&\\
 & $0.43^{+0.04}_{-0.04}$&$-1.24^{+0.22}_{-0.20}$& \\ 
 SNIa(new Gold)+CMB&&& $15.21 ^{+3.66}_{-3.57}$ \\
&$0.43^{+0.09}_{-0.08}$&$-1.24^{+0.39}_{-0.47}$&
 \\
 &&&\\ \hline
 &&& \\
SNIa(new Gold)+& $0.33^{+0.02}_{-0.03}$&$-1.92^{+0.18}_{-0.25}$&
 \\
 CMB+SDSS&&& $15.62^{+3.14}_{-3.91}$ \\
&$0.33^{+0.05}_{-0.05}$&$-1.92^{+0.37}_{-0.49}$ & \\
 &&&\\  
 \hline &&&\\

  & $0.16^{+0.28}_{-0.16}$&$-4.70^{+3.43}$ &  \\ 
 SNIa (SNLS)&&& $12.64 ^{+8.74}$\\
 & $0.16^{+0.35}_{-0.16}$&$-4.70^{+3.73}$ &
\\ &&&\\ \hline
&&&\\
& $0.33^{+0.05}_{-0.05}$&$-1.91^{+0.38}_{-0.53}$ & \\
SNIa(SNLS)+CMB&&& $14.30^{+4.49}_{-4.86}$\\
&$0.33^{+0.10}_{-0.09}$&$-1.91^{+0.69}_{-1.23}$&
 \\
 &&&\\ \hline
 &&& \\
SNIa(SNLS)+& $0.28^{+0.02}_{-0.02}$&$-2.47^{+0.27}_{-0.30}$&
 \\
 CMB+SDSS&&&$15.23 ^{+4.04}_{-4.07}$  \\
&$0.28^{+0.05}_{-0.04}$&$-2.47^{+0.52}_{-0.64}$ & \\
 &&&\\
\end{tabular}
\end{center}
\end{table}
\begin{table}[t]
\begin{center}
\caption{\label{tab3} The best fit values for the parameters of the
model using SNIa from new Gold sample and SNLS data, SNIa+CMB and
SNIa+CMB+SDSS experiments at one and two $\sigma$ confidence level.
Here we suppose $w=0.0$. }
\begin{tabular}{|c|c|c|c|}
Observation & $\Omega_m$ & $\Omega_{\cal{C}}$ & $w_r$
\\ \hline
  &&& \\
 & $0.16^{+0.84}_{-0.03}$&$0.20_{-0.19}^{+0.04}$&$-5.80^{+4.80}$  \\ 
 SNIa(new Gold)&&&\\
 & $0.16^{+0.84}_{-0.08}$&$0.20_{-0.19}^{+0.08}$&$-5.80^{+4.90}$
\\ &&&\\ \hline
&&&\\
SNIa(new Gold)+CMB & $0.53^{+0.10}_{-0.09}$&$0.03^{+0.02}_{-0.02}$& $-1.10 ^{+0.20}_{-0.20}$ \\ 
 &&& \\
&$0.53^{+0.16}_{-0.19}$&$0.03^{+0.03}_{-0.03}$&$-1.10
^{+0.40}_{-0.70}$
 \\
 &&&\\ \hline
 &&& \\
SNIa(new Gold)+&
$0.31^{+0.02}_{-0.02}$&$0.05^{+0.01}_{-0.01}$&$-1.40
^{+0.20}_{-0.20}$
 \\
 CMB+SDSS&&&  \\
&$0.31^{+0.02}_{-0.02}$&$0.05^{+0.01}_{-0.01}$ &$-1.40 ^{+0.20}_{-0.20}$ \\
 &&&\\  
 \hline &&&\\

  & $0.39^{+0.47}_{-0.26}$&$0.054^{+0.056}_{-0.053}$ & $-1.80 ^{+0.70}$  \\ 
 SNIa (SNLS)&&&\\
 & $0.39^{+0.61}_{-0.30}$&$0.054^{+0.086}_{-0.053}$ & $-1.80 ^{+0.80}$
\\ &&&\\ \hline
&&&\\
SNIa(SNLS)+CMB& $0.42^{+0.09}_{-0.23}$&$0.027^{+0.013}_{-0.023}$ & $-1.60 ^{+0.30}_{-1.05}$ \\
&&& \\
&$0.42^{+0.12}_{-0.33}$&$0.027^{+0.018}_{-0.026}$ & $-1.60
^{+0.45}_{-5.00}$
 \\
 &&&\\ \hline
 &&& \\
SNIa(SNLS)+& $0.28^{+0.03}_{-0.02}$&$0.037^{+0.003}_{-0.004}$ &
$-2.05 ^{+0.15}_{-0.15}$
 \\
 CMB+SDSS&&&  \\
&$0.28^{+0.05}_{-0.04}$&$0.037^{+0.008}_{-0.008}$ &
$-2.05 ^{+0.35}_{-0.35}$ \\
 &&&\\
\end{tabular}
\end{center}
\end{table}
\begin{table}[t]
\begin{center}
\caption{\label{tab4} The best values for the curvature of a thick
brane model with the corresponding age for the universe from fitting
with SNIa from new Gold sample and SNLS data, SNIa+CMB and
SNIa+CMB+SDSS experiments at one and two $\sigma$ confidence level.
Here we suppose $w=0.0$. }
\begin{tabular}{|c|c|c|}
Observation & $\Omega_k$ & Age (Gyr)
\\ \hline
   &&\\
    SNIa(new Gold)&$-0.28^{+5.79}_{-1.62}$&$12.54^{+8.85}$\\
&&\\ \hline
&&\\
SNIa(new Gold)+CMB & $+0.14^{+0.24}_{-0.22}$&$15.31^{+1.06}_{-1.81}$ \\ 
 && \\
 \hline
 && \\
SNIa(new Gold)+&$+0.21^{+0.08}_{-0.08}$&$14.72^{+0.43}_{-0.48}$
 \\
 CMB+SDSS&&  \\
  \hline &&\\
  SNIa (SNLS)&$-0.14^{+1.34}$&$13.50^{+2.95}$\\
 && \\ \hline
&&\\
SNIa(SNLS)+CMB& $-0.10^{+0.27}_{-0.74}$&$14.66^{+1.02}_{-4.61}$ \\
&&\\ \hline
 && \\
SNIa(SNLS)+& $+0.11^{+0.10}_{-0.07}$&$14.18^{+0.26}_{-0.29}$ \\
 CMB+SDSS&&  \\
&&\\
\end{tabular}
\end{center}
\end{table}
\begin{table}
\begin{center}
\caption{\label{tab5} The value of $\tau$ for three high redshift
objects, using the parameters of the model derived from fitting with
the observations. Here we imagined flat universe and $w=0.0$. }
\begin{tabular}{|c|c|c|c|}
  Observation & LBDS &LBDS  & APM  \\
& $53$W$069$&$53$W$091$& $08279+5255$ \\
  & $z=1.43$&$z=1.55$& $z=3.91$  \\ \hline
  &&&\\
SNIa (new Gold)& $0.35^{+0.13}_{-0.02}$ & $0.36^{+0.13}_{-0.02}$& $0.16^{+0.07}_{-0.01}$ \\
&&&\\
 \hline

&&&\\
SNIa(new Gold)+CMB &$1.27^{+0.54}_{-0.52}$&$1.36^{+0.58}_{-0.57}$&$0.75^{+0.41}_{-0.37}$ \\
 +SDSS& && \\
\hline
&&&\\
SNIa(new Gold)+CMB & $1.33^{+0.48}_{-0.53}$&$1.41^{+0.52}_{-0.57}$&$0.75^{+0.36}_{-0.38}$ \\
 +SDSS+LSS& && \\ \hline
&&&\\
 SNIa (SNLS)& $ 0.85^{+0.64}_{-0.85}$ & $0.89^{+0.67}_{-0.89}$& $0.42^{+0.33}_{-0.42}$ \\ &&&\\ \hline
&&&\\
SNIa(SNLS)+CMB & $1.19^{+0.61}_{-0.64}$&$1.27^{+0.65}_{-0.69}$&$0.67^{+0.41}_{-0.40}$ \\
 +SDSS& && \\
\hline
&&&\\
SNIa(SNLS)+CMB & $1.36^{+0.62}_{-0.62}$&$1.45^{+0.67}_{-0.67}$&$0.78^{+0.45}_{-0.42}$ \\
 +SDSS+LSS& && \\
\end{tabular}
\end{center}
\end{table}

\begin{table}
\begin{center}
\caption{\label{tab6} The value of $\tau$ for three high redshift
objects, using the parameters of the model derived from fitting with
the observations. Here we chose $w=0.0$. }
\begin{tabular}{|c|c|c|c|}
  Observation & LBDS &LBDS  & APM  \\
& $53$W$069$&$53$W$091$& $08279+5255$ \\
  & $z=1.43$&$z=1.55$& $z=3.91$  \\ \hline
  &&&\\
SNIa (new Gold)& $0.75^{+0.44}_{-0.75}$ & $0.78^{+0.47}_{-0.78}$& $0.35^{+0.27}_{-0.35}$ \\
&&&\\
 \hline

&&&\\
SNIa(new Gold)+CMB &$1.37^{+0.23}_{-0.41}$&$1.46^{+0.25}_{-0.46}$&$0.83^{+0.24}_{-0.45}$ \\
 +SDSS& && \\
\hline
&&&\\
SNIa(new Gold)+CMB & $1.13^{+0.07}_{-0.08}$&$1.19^{+0.07}_{-0.09}$&$0.65^{+0.14}_{-0.07}$ \\
 +SDSS+LSS& && \\ \hline
&&&\\
 SNIa (SNLS)& $ 1.10^{+0.34}_{-1.10}$ & $1.16^{+0.43}_{-1.16}$& $0.58^{+0.29}_{-0.58}$ \\ &&&\\ \hline
&&&\\
SNIa(SNLS)+CMB & $1.35^{+0.25}_{-1.31}$&$1.44^{+0.27}_{-1.42}$&$0.80^{+0.22}_{-0.80}$ \\
 +SDSS& && \\
\hline
&&&\\
SNIa(SNLS)+CMB & $1.17^{+0.05}_{-0.05}$&$1.24^{+0.06}_{-0.06}$&$0.67^{+0.14}_{-0.04}$ \\
 +SDSS+LSS& && \\
\end{tabular}
\end{center}
\end{table}

\section{Age of Universe}

\begin{figure}[t]
\epsfxsize=8truecm\epsfbox{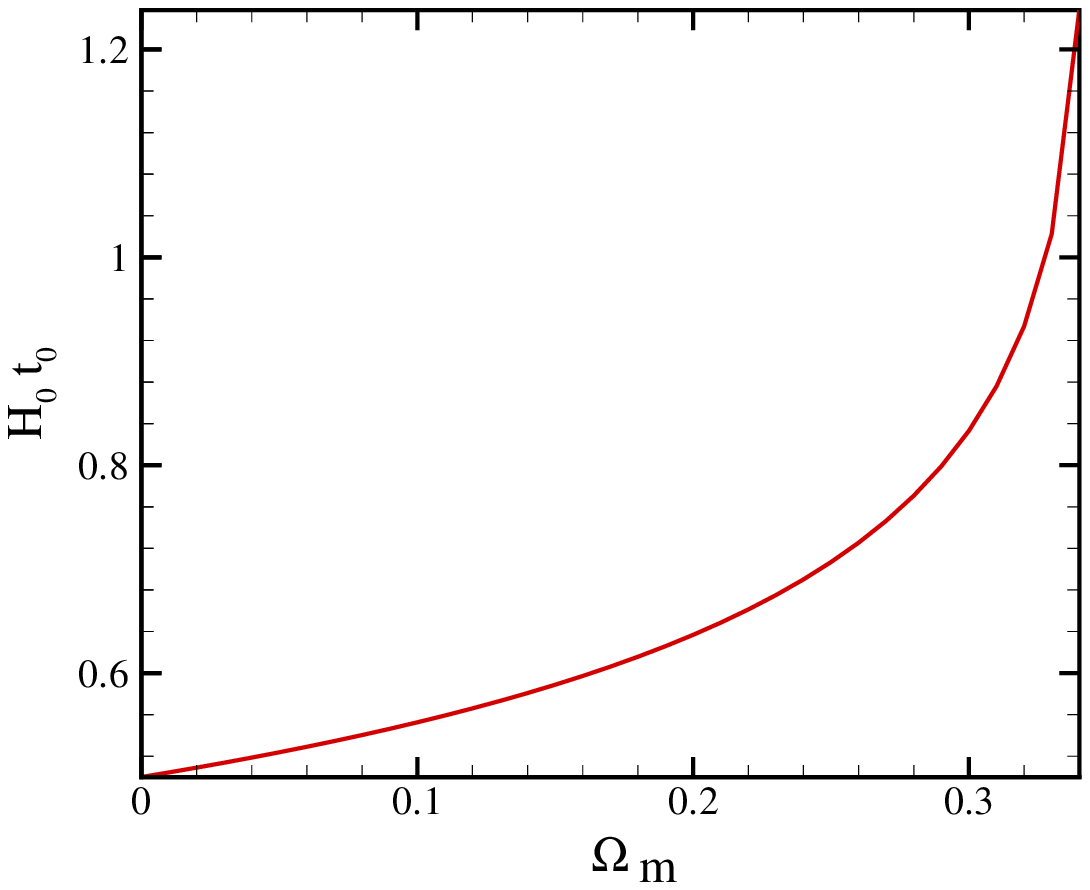} \narrowtext
\epsfxsize=8truecm\epsfbox{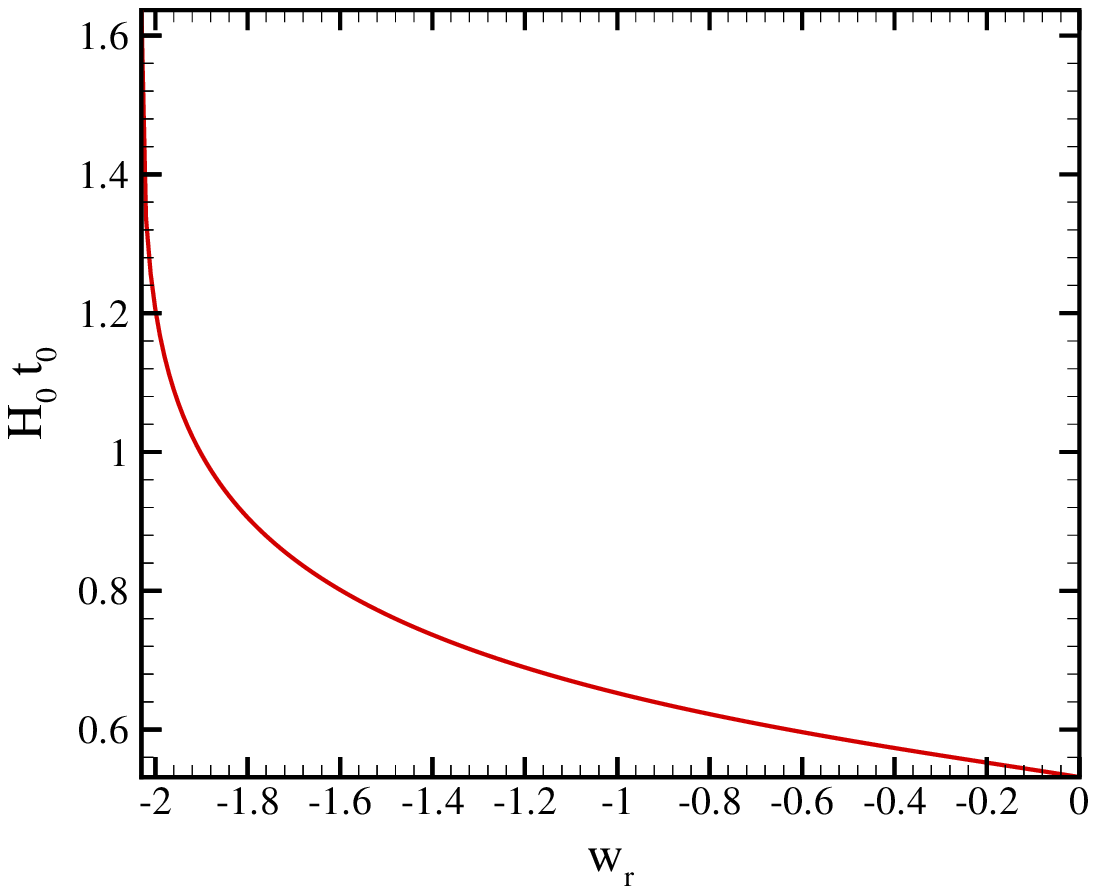}\narrowtext\caption{ $H_0t_0$
(age of universe times the Hubble constant at the present time) as a
function of $\Omega_m$ (upper panel) for a flat universe dominated
by cold dark matter and typical value of $w_r=-1.92$. Increasing
$\Omega_m$ gives a longer age for the universe. Lower panel shows
the same function versus $w_r$ for the case $w=0.0$, $\Omega_m=0.33$
and flat universe.} \label{age1}
 \end{figure}

\begin{figure}[t]
\epsfxsize=8truecm\epsfbox{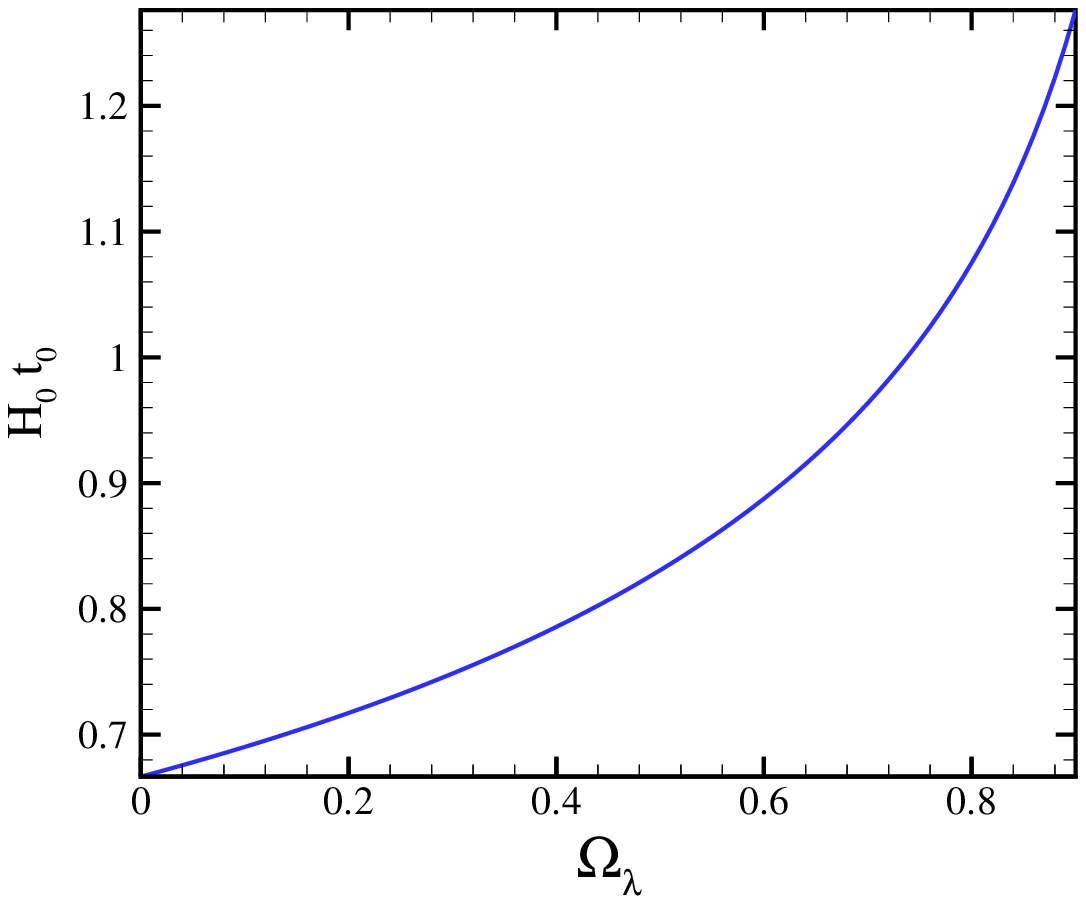} \narrowtext
\epsfxsize=8truecm\epsfbox{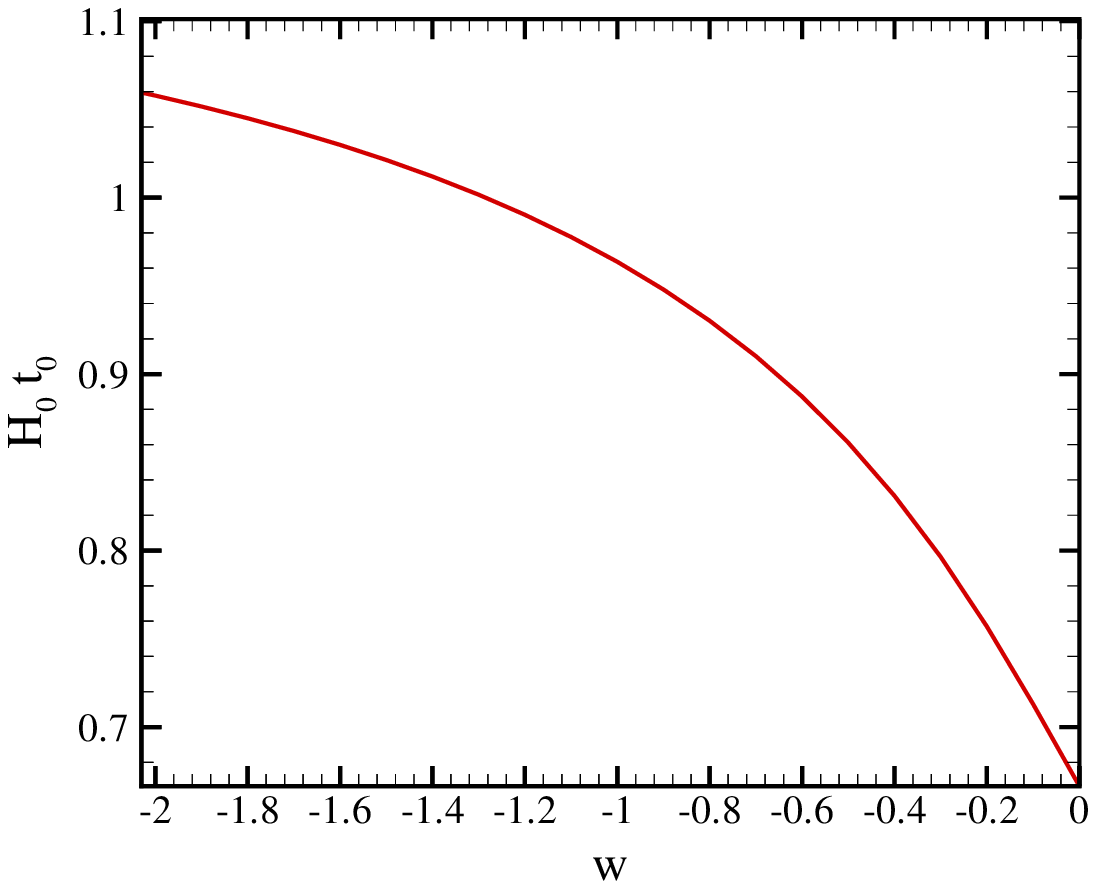} \narrowtext
 \caption{
$H_0t_0$ (age of universe times the Hubble constant at the present
time) versus $\Omega_{\lambda}$ (upper panel) in the $\Lambda$CDM
model for the case $\Omega_k=0.0$ and $w=-1.0$. Lower panel shows
$H_0t_0$ as a function of present equation of state, $w$, in the
$\Lambda$CDM model with the typical values $\Omega_m=0.30$ and
$\Omega_{\lambda}=0.70$.} \label{agel}
\end{figure}

The age of universe integrated from the big bang up to now in terms
of free parameters of thick brane model is given by:

\begin{eqnarray}\label{age}
&&t_0(\Omega_m,\Omega_{\cal{C}},w_r,w) = \int_0^{t_0}\,dt \nonumber\\
&&={1 \over H_0\sqrt{|\Omega_k|}}\, {\cal F} \left(
\sqrt{|\Omega_k|}\int_0^\infty {dz'H_0\over (1+z') H(z')} \right)
\end{eqnarray}
Figure (\ref{age1}) shows the dependence of $H_0t_0$ (Hubble
parameter times the age of universe) on $\Omega_m$ and $w_r$ for a
flat universe. Obviously increasing $\Omega_m$ and $w_r$ result in a
longer and shorter age for the universe, respectively. As a matter
of fact, according to the equation (\ref{hub}),  $\Omega_m |w_r|$
behaves as dark energy in the $\Lambda$CDM scenario and $w_r$ has
the same role as $w$ in the $\Lambda$CDM (see figures (\ref{agel})).

 The "age crisis" is one the main reasons
of the acceleration phase of the universe. The problem is that the
universe's age in the Cold Dark Matter (CDM) universe is less than
the age of old stars in it. Studies on the old stars
\cite{carretta00} suggest an age of $13^{+4}_{-2}$ Gyr for the
universe. Richer et. al. \cite{richer02} and Hasen et. al.
\cite{hansen02} also proposed an age of $12.7\pm0.7$ Gyr, using the
white dwarf cooling sequence method (for full review of the cosmic
age see \cite{spe03}). To do another consistency test, we compare
the age of universe derived from this model with the age of old
stars and Old High Redshift Galaxies (OHRG) in various redshifts.
Table \ref{tab4} shows that age of the universe from the combined
analysis of SNIa$+$CMB$+$SDSS is $14.72_{-0.48}^{+0.43}$ Gyr and
$14.18_{-0.29}^{+0.26}$ Gyr for new Gold sample and SNLS data,
respectively, while $\Lambda$CDM implies $13.7\pm0.2$Gyr
\cite{spe03}. These values are in agreement with the age of old
stars \cite{carretta00}.

Here we consider three OHRG for comparison with the thick brane
model, namely the LBDS $53$W$091$, a $3.5$-Gyr old radio galaxy at
$z=1.55$ \cite{dunlop96}, the LBDS $53$W$069$ a $4.0$-Gyr old radio
galaxy at $z=1.43$ \cite{dunlop99} and a quasar, APM $08279+5255$ at
$z=3.91$ with an age of $t=2.1_{-0.1}^{+0.9}$Gyr \cite{hasinger02}.
The latter has once again led to the "age crisis". An interesting
point about this quasar is that it cannot be accommodated in the
$\Lambda$CDM model \cite{jan06}. In order to quantify the
age-consistency test we introduce the expression $\tau$ as:
\begin{equation}
 \tau=\frac{t(z;\Omega_m,\Omega_{\cal{C}},w_r,w)}{t_{obs}} = \frac{t(z;\Omega_m,\Omega_{\cal{C}},w_r,w)H_0}{t_{obs}H_0}
\end{equation}
where $t(z)$ is the age of universe, obtained from the equation
(\ref{age}) and $t_{obs}$ is an estimation for the age of old
cosmological object. In order to have a compatible age for the
universe we should have $\tau>1$. Tables \ref{tab5} and \ref{tab6}
report the value of $\tau$ for three mentioned OHRG with various
observations. We see that the parameters of thick brane model from
the combined observations provide a compatible age for the universe,
compared to the age of old objects, also in addition SNLS data
result in a shorter age for the universe. Once again for the thick
brane model, APM $08279+5255$ at $z=3.91$ has a longer age than the
universe but gives better results than most cosmological models
investigated before \cite{sa1,sa2,jan06}.

\section{conclusions and discussions}
From observational point of view, it has been possible to compare
theoretical model with the observational results.

We explored the consistency of a thick codimension $1$ brane model
with the implication of up-to-date luminosity of supernova type Ia
observed by two independent groups, new Gold sample and SNLS data
set, acoustic peak in the cosmic microwave background anisotropy
power spectrum and baryon acoustic oscillation measured by Sloan
Digital Sky Survey.

 In this scenario, universe is supposed to be dominated
by pressureless cold dark matter, which penetrates to the extra
dimension, this leads to an acceleration epoch for the universe.

In this work we have been interested in matter dominated era for the
universe. So we imagined $w=0.0$ as a prior through this paper. The
best parameters obtained from the fitting with the new Gold sample
data combined with CMB and SDSS observations are:
$\Omega_m=0.31_{-0.02}^{+0.02}$,
$\Omega_{\cal{C}}=0.05_{-0.01}^{+0.01}$ and
$w_r=-1.40^{+0.20}_{-0.20}$ at $1\sigma$ confidence level states
spatially open universe with  $\Omega_k=+0.21_{-0.08}^{+0.08}$. SNLS
SNIa$+$CMB$+$SDSS give: $\Omega_m=0.28_{-0.02}^{+0.03}$,
$\Omega_{\cal{C}}=0.037_{-0.004}^{+0.003}$ and
$w_r=-2.05^{+0.15}_{-0.15}$ at $1\sigma$ confidence level
demonstrate $\Omega_k=+0.11_{-0.07}^{+0.10}$. The well-known
$\Lambda$CDM model implying $-0.06\le\Omega_k\le+0.02$ \cite{spe03}
and  some other interesting models such as Dvali-Gabadadze-Porrati
(DGP) which states $\Omega_k=0.01^{+0.09}_{-0.09}$ and
$\Omega_k=0.01^{+0.04}_{-0.04}$ using Gold sample and SNLS data,
respectively \cite{0603632v2,sadgp}, show a contradiction with our
results. In fact having a spatially open universe is ruled out in
many models comparing to the observations while in this thick brane
model we find out it is not possible to have a spatially flat
universe, according to the recent observational tests. The value of
$w_r$ given by observational constraints is negative. This shows
that instead of dark energy to accelerate the universe, we have a
strange effect of matter through the extra dimension with negative
pressure.

We also performed the age test, comparing the age of old stars and
old high redshift galaxies with the age derived from this model.
From the best fit parameters of the model using new Gold sample and
SNLS SNIa, respectively,  we obtained an age of
$14.72_{-0.48}^{+0.43}$ Gyr and $14.18_{-0.29}^{+0.26}$ Gyr, for the
universe. These results are in agreement with the age of the old
stars. The age of universe in this model is larger than what is
given in the other models \cite{spe03,sa1,sa2,sadgp}

 To check the age crisis in this model we chose two high
redshift radio galaxies at $z=1.55$ and $z=1.43$ with a quasar at
$z=3.91$. Two first objects were consistent with the age of
universe, i.e., they were younger than the universe while the third
one was not but gave better result than $\Lambda$CDM and a class of
Quintessence model \cite{sa1,sa2}.


\end{document}